\documentclass[sigconf,natbib]{acmart}
\usepackage{amsmath}
\usepackage{graphicx}
\usepackage{xpatch}
\xpretocmd{\eqref}{Eq.~}{}{}

\def\figscale{.48}

\newcommand{\sub}[1]{\textit{\footnotesize{#1}}}
\newcommand{\boldsec}[1]{\ensuremath{\texorpdfstring{\boldsymbol{#1}}{#1}}}
\newcommand{\ind}[1]{\mathbb{I}\!\left[#1\right]}

\newcommand{\permS}{\ensuremath{\tilde{S}}}
\newcommand{\permL}{\ensuremath{\tilde{L}}}
\newcommand{\PS}{\ensuremath{\mathcal{P}_S}}
\newcommand{\PL}{\ensuremath{\mathcal{P}_L}}

\newcommand{\contS}{\ensuremath{c_{e,S|d}}}
\newcommand{\contL}{\ensuremath{c_{e,L|d}}}

\newcommand{\topS}{\ensuremath{t_{e,S}}}
\newcommand{\botS}{\ensuremath{b_{e,S}}}

\newcommand{\posS}{\ensuremath{S_{e}^{-1}}}
\newcommand{\posL}{\ensuremath{L_{e}^{-1}}}
\newcommand{\posSt}{\ensuremath{\tilde{S}_{e}^{-1}}}
\newcommand{\posLt}{\ensuremath{\tilde{L}_{e}^{-1}}}

\copyrightyear{2024}
\acmYear{2024}
\setcopyright{rightsretained}
\acmConference[SIGIR '24]{Proceedings of the 47th International ACM SIGIR Conference on Research and Development in Information Retrieval}{July 14--18, 2024}{Washington, DC, USA} \acmBooktitle{Proceedings of the 47th International ACM SIGIR Conference on Research and Development in Information Retrieval (SIGIR '24), July 14--18, 2024, Washington, DC, USA} \acmDOI{10.1145/3626772.3657700} \acmISBN{979-8-4007-0431-4/24/07}

\makeatletter
\gdef\@copyrightpermission{
	\begin{minipage}{0.3\columnwidth}
		\href{https://creativecommons.org/licenses/by/4.0/}{\includegraphics[width=0.90\textwidth]{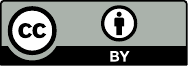}}
	\end{minipage}\hfill
	\begin{minipage}{0.7\columnwidth}
		\href{https://creativecommons.org/licenses/by/4.0/}{This work is licensed under a Creative Commons Attribution International 4.0 License.}
	\end{minipage}
	\vspace{5pt}
}
\makeatother

\begin{document}
\title{The Treatment of Ties in Rank-Biased Overlap}

\author{Matteo Corsi}
\affiliation{%
	\institution{Delft University of Technology}
	\city{Delft}\country{The Netherlands}
}
\email{m.corsi@tudelft.nl}

\author{Juli\'an Urbano}
\affiliation{%
	\institution{Delft University of Technology}
	\city{Delft}\country{The Netherlands}
}
\email{j.urbano@tudelft.nl}

\begin{abstract}
Rank-Biased Overlap ($RBO$) is a similarity measure for indefinite rankings: it is top-weighted, and can be computed when only a prefix of the rankings is known or when they have only some items in common. It is widely used for instance to analyze differences between search engines by comparing the rankings of documents they retrieve for the same queries. In these situations, though, it is very frequent to find tied documents that have the same score.
Unfortunately, the treatment of ties in $RBO$ remains superficial and incomplete, in the sense that it is not clear how to calculate it from the ranking prefixes only. In addition, the existing way of dealing with ties is very different from the one traditionally followed in the field of Statistics, most notably found in rank correlation coefficients such as Kendall’s and Spearman’s. 
In this paper we propose a generalized formulation for $RBO$ to handle ties, thanks to which we complete the original definitions by showing how to perform prefix evaluation. We also use it to fully develop two variants that align with the ones found in the Statistics literature: one when there is a reference ranking to compare to, and one when there is not. Overall, these three variants provide researchers with flexibility when comparing rankings with $RBO$, by clearly determining what ties mean, and how they should be treated.
Finally, using both synthetic and TREC data, we demonstrate the use of these new tie-aware $RBO$ measures. We show that the scores may differ substantially from the original tie-unaware $RBO$ measure, where ties had to be broken at random or by arbitrary criteria such as by document ID. Overall, these results evidence the need for a proper account of ties in rank similarity measures such as $RBO$.
\end{abstract}



\maketitle\sloppy

\section{Introduction}\label{sec:intro}

Rankings are part of our everyday lives: music albums are ranked by sales, universities are ranked by research output, cities are ranked by livability, etc. In Information Retrieval (IR) and Recommender Systems (RecSys), rankings are essential: search engines rank documents by likelihood of relevance to a query, and recommenders rank for instance books by likelihood of purchase.
But rankings can be made following alternative criteria, such as music albums by replays, universities by alumni success, cities by pollution, etc. One way to understand the differences and commonalities of ranking criteria is to compare the rankings they produce. In IR this happens when we compare the rankings of documents returned by different systems, topics that documents are estimated to fit to, terms for query expansion, or rankings of systems sorted by different evaluation metrics or different relevance assessors.

Comparing rankings requires a rank similarity measure. Some of the most well-known examples are $\tau$ by \citet{kendall1938new}, $\rho$ by \citet{spearman1904proof}, and $D$ by \citet{kolmogorov1933sulla}, while others developed in the context of IR and related disciplines are $\tau_*$ by \citet{melucci2007rank}, $d_\text{rank}$ by \citet{carterette2009rank}, $\tau_{ap}$ by \citet{yilmaz2008new}, $K^*$ and $F'^*$ by \citet{kumar2010generalized}, and $\tau_w$ by \citet{vigna2015weighted}.
While some of these are top-weighted and thus assign more importance to similarities at the top of rankings than at the bottom, none of them can compare non-conjoint rankings that have only some items in common. This is very often the case in IR and RecSys when comparing the results from search engines that have different indexes, recommenders that have different catalogs, or simply cases where the rankings are truncated after a certain depth.
The problem of rank similarity under non-conjointness has received far less attention, with works inspired by Spearman's footrule \citep{fagin2003comparing,barillan2006methods}, the Hoeffding distance \citep{sun2010visualizing}, and even IR metrics \citep{buckley2004topic,tan2015family}. Most notably, \citet{Webber2010} proposed Rank-Biased Overlap ($RBO$), which has become popular in IR research for example to compare search engine results \citep{cardoso2011google, pochat2019tranco}, measure topic similarity \citep{abdelrazek2022topic, mantyla2018lda}, assess consistency of systems to query variations \citep{bailey2017retrieval}, or compare rankings of documents in general \citep{clarke2020offline, moffat2017user, zendel2019information,salavati2017bridgerank}. Beyond IR, it is also used for example in RecSys  \citep{Vrijenhoek2021recommenders, Canamares2020offline}, Network Science \citep{rajeh2020interplay, ma2020lgiem} and Neuroscience \citep{steurangel2022hypomap, buch2023molecular}.

$RBO$ is top-weighted, and it handles non-conjointness as well as incomplete rankings, even of different lengths. Incomplete rankings appear for instance after truncation, because their very top-weighted nature implies that, after a sufficiently deep rank, what items appear next is negligible. For example, a search engine may return only the top 20 documents in response to a query, a recommender may suggest only the top 5 items, and a typical TREC run consists of only the top 1,000 documents per topic. This means that rankings may actually consist of a \emph{seen} part or prefix, and an \emph{unseen} part that extends further, potentially up to infinity. Therefore, $RBO$ scores have to be computed from a prefix only, ideally accompanied by some quantification of the uncertainty due to the unseen items. \citet{Webber2010} showed how to compute upper and lower bounds for this uncertainty, as well as a point estimate, called $RBO_\sub{EXT}$. This is the score typically reported when using $RBO$.

\subsection{The Problem of Tied Items}\label{ssec:ties}

\begin{table}[!t]
\caption{Summary statistics of the adhoc runs in the last editions of TREC Web. Fifty topics were used in all cases.}\label{tab:trec-summary}
\centering\small\begin{tabular}{|r|rrrrrr|}
\hline
& &  & Runs & Rankings & Docs & Avg. tie \\
Year & Groups & Runs & with ties & with ties & tied & group size \\ \hline
2009 & 25 & 71 & 90\% & 88\% & 22\% & 19.3 \\
2010 & 21 & 56 & 96\% & 87\% & 30\% & 17.7 \\
2011 & 14 & 37 & 89\% & 79\% & 27\% & 3.6 \\
2012 & 11 & 27 & 63\% & 63\% & 20\% & 2.8 \\
2013 & 14 & 34 & 62\% & 57\% & 35\% & 26.6 \\
2014 & 10 & 30 & 60\% & 60\% & 24\% & 5.1 \\ \hline
Avg. & 16 & 43 & 77\% & 72\% & 26\% & 12.5 \\ \hline
\end{tabular}
\end{table}

Rankings may very well contain tied items. For example, systems with the same $P@10$ score, or documents with the same retrieval score for a query. The latter is a well-known issue in IR research. For instance, Table~\ref{tab:trec-summary} presents summary statistics from all the adhoc runs in TREC Web between 2009 and 2014, showing that 77\% of the runs contained ties in 72\% of the rankings, with 26\% of the documents tied in groups of 12.5 documents on average.
This begs the question: how should these documents be ranked when evaluating effectiveness? Different approaches to this problem, as well as their impact, have been studied for example by \citet{raghavan1989critical,mcsherry2008computing,cabanac2010tie,lin2019impact}.

\begin{figure}[!t]
\includegraphics[scale=\figscale]{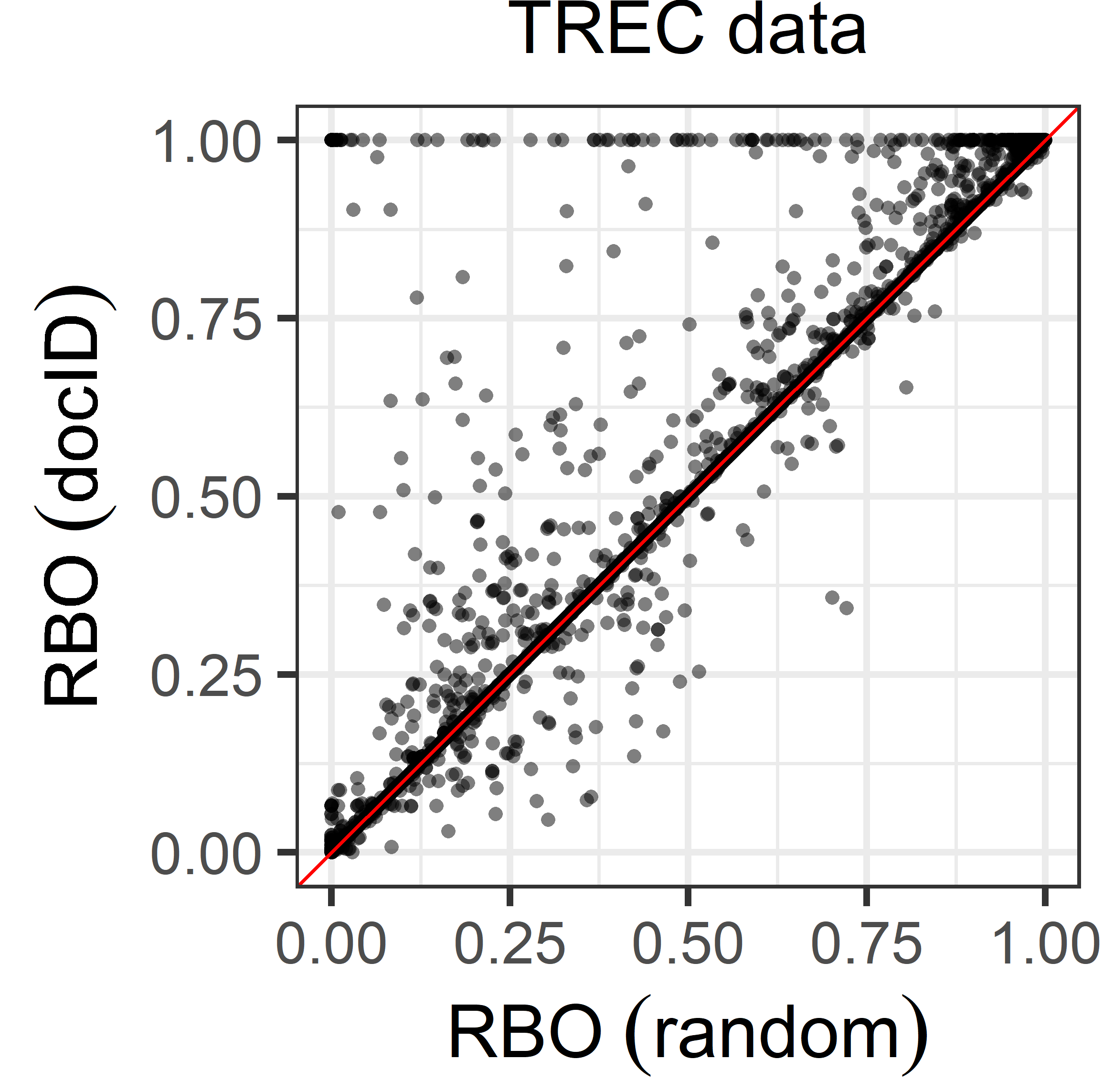}~~~~
\includegraphics[scale=\figscale]{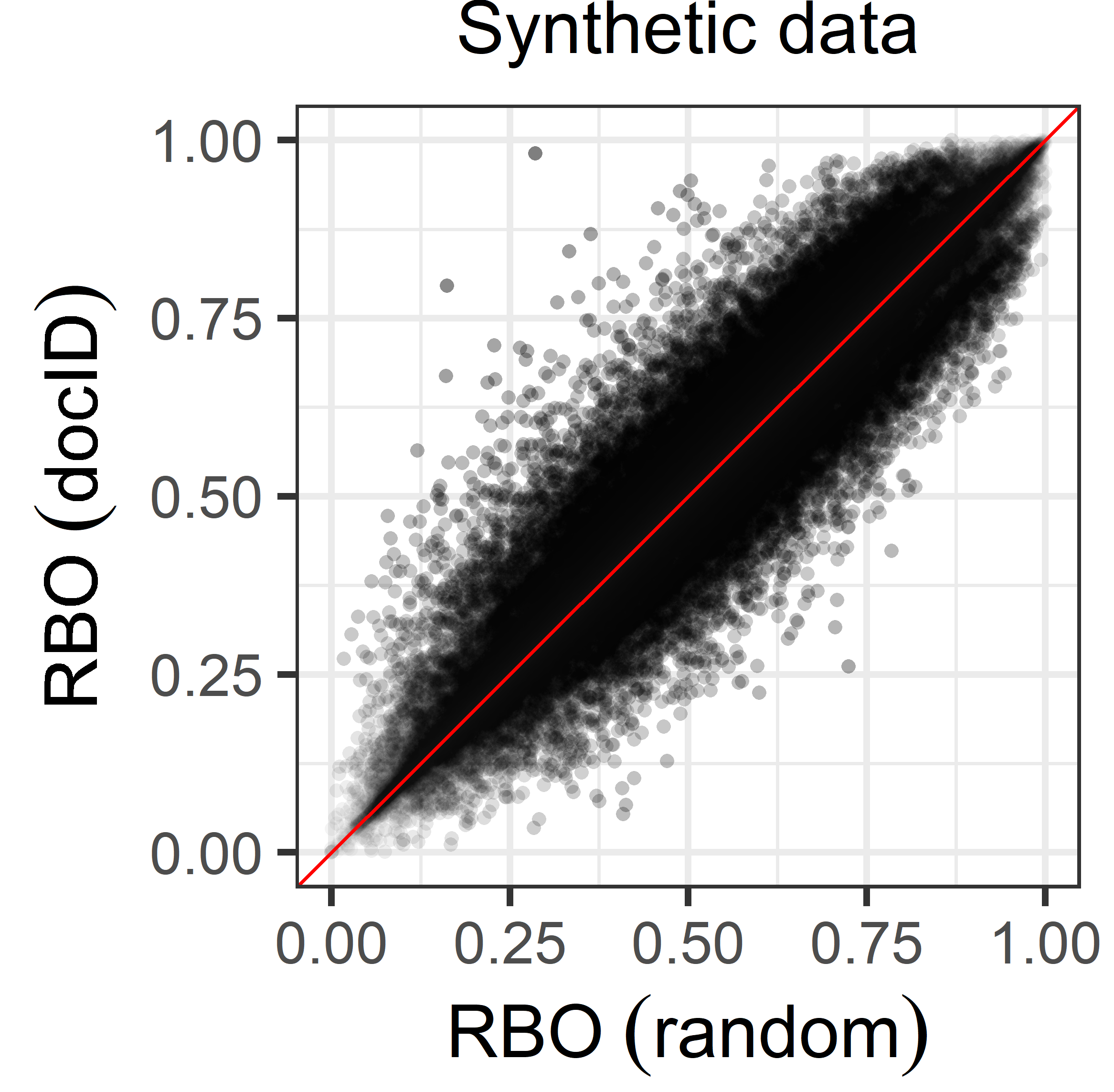}
\caption{Differences in $\boldsec{RBO (p=0.9)}$ when breaking ties by doc ID or at random, for TREC and synthetic data}\label{fig:breaker}
\end{figure}

A similar question may be asked about rank similarity measures such as $RBO$: how should tied items be handled? The first approach is to break the ties, essentially ignoring them. For example, ties may be broken at random, but this would introduce noise. Other, rather arbitrary criteria may be used to break ties, such as by document ID; this is the approach implemented in \texttt{trec\_eval}, but it has its own issues~\citep{cabanac2010tie,lin2019impact}. For $RBO$ specifically, it inflates similarity scores because tied documents would artificially appear in the same order in both rankings. We illustrate this in Figure~\ref{fig:breaker} with the $RBO$ scores computed between pairs of the TREC Web runs summarized in Table~\ref{tab:trec-summary}, as well as synthetic rankings (details are presented in Section~\ref{sec:exp}). As expected, these tie-breaking strategies generally lead to different $RBO$ scores, but some differences are strikingly large, specially in the TREC data. Differences are larger than normal reporting fidelity (i.e. 2 decimal digits) in 10\% of the TREC cases, and 66\% of the synthetic ones. In addition, it is clear that breaking ties by document ID does inflate scores.

The second approach is to explicitly handle the ties, in a principled way, in $RBO$ itself. \citet{Webber2010} followed this line to motivate a tie-aware variant by assuming that all items tied between ranks $n$ and $m$ occur at the same rank $n$. In essence, they assumed that tied items \emph{really} occur at the same rank.
This contrasts with the view typically taken in the Statistics literature, where a tie represents \emph{uncertainty}. Specifically, an integral ranking is assumed to exist, with a strict order between all items, and a tie really represents that it is not known, for whatever reason, which item goes first. This happens when their underlying scores are the same within reporting fidelity, or when they represent observations that are too noisy to discern the actual order. In essence, a tie here represents a loss of information.
This notion of ties for rank similarity dates back to more than a century ago, to the work of \citet{student1921experimental} for adapting Spearman's $\rho$ to handle ties. It was later popularized, most notably by Kendall~\citep{kendall1948rank,kendall1945treatment}, when similarly adapting his $\tau$ coefficient, which resulted in two tie-aware variants: $\tau_a$ and $\tau_b$.

The $a$-variant was proposed for cases where one ranking represents a reference and the goal is to compute the \emph{accuracy} of the other ranking with respect to this reference. To this end, $\tau_a$ was precisely defined as the expected value of $\tau$ when breaking ties at random.
The $b$-variant was proposed for cases where there is no reference and one wants to calculate the \emph{agreement} between the two rankings. To this end, $\tau_b$ corrects the measured similarity by the amount of information lost due to ties.
\citet{vigna2015weighted} followed the second approach to define his $\tau_w$ coefficient, albeit without explicit mention of it. More recently, \citet{urbano2017treatment} followed both approaches to define tie-aware variants of \citeauthor{yilmaz2008new}'s $\tau_{ap}$.

\subsection{Contributions}\label{ssec:contributions}

\citet{Webber2010} tackled the issue of ties in $RBO$ only in passing: as will be detailed in Sections~\ref{ssec:ties-a} and \ref{sec:prefix}, it is unclear how to calculate $RBO_\sub{EXT}$ and its bounds when dealing with ties. This is very well illustrated by one of the popular implementations available online,\footnote{\url{https://github.com/dlukes/rbo}} where authors argue that the equations in \citep{Webber2010} need modifications in order to handle ties, but it turns out that these modifications make the results incorrect when rankings do not have ties.

In this paper we deal with the problem of explicitly treating ties in Rank-Biased Overlap. Specifically:
\begin{enumerate}
	\item We show how to compute both $a$- and $b$-variants of $RBO$.
	\item We develop a formulation for $RBO$ that generalizes all three variants and allows us to derive the point estimate $RBO_\sub{EXT}$ and the bounds (i.e., $RBO_\sub{MIN}$, $RBO_\sub{MAX}$ and thus $RBO_\sub{RES}$). 
	\item Using both TREC data and synthetic data we illustrate the differences among variants, as well as the importance of following a principled approach to deal with ties, as opposed to arbitrarily breaking them and computing bare $RBO$.
	\item We provide a full implementation of all coefficients,\footnote{\url{https://github.com/julian-urbano/sigir2024-rbo}} as well as guidelines for when to use each (see Section~\ref{sec:conclusions}).
\end{enumerate}

In summary, we contribute the theoretical underpinnings for a principled treatment of ties in $RBO$, providing complete formulations and implementations of three variants that align with different notions of ties, namely $RBO^{\!w}, RBO^a$ and $RBO^b$.
\section{Rank-Biased Overlap}\label{sec:rbo}

\begin{table}[!t]
\caption{Summary of notation. See Figure~\ref{fig:example} for examples.}\label{tab:notation}
\small\begin{tabular}{|l|l|} 
\hline
$S, L$ & Rankings of lengths $s$ and $l$, where $s\leq l$. \\
$S_d,~S_e^{-1}$ & Item at rank $d$ and rank of item $e$, in $S$. \\
$S_{n:m}$ & Set of items from rank $n$ to $m$ in $S$. \\	
$\Omega= \{S \cup L\}$ & Set of all items seen in $S$ or $L$. \\
\hline
$d$ & Evaluation depth for computing agreement. \\
$X_d, A_d$ & Overlap and agreement at depth $d$.\\
$p$ & Persistence parameter of $RBO$.\\
\hline	
$\topS,~\botS$ & Top and bottom ranks of $e$'s tie group in $S$. \\
$c_{e,S|d}$ & Contribution of item $e$ in $S$ given depth $d$. \\
\hline
Inactive item & One that is surely below $d$ (i.e. $d < t_e$). \\
Active item & One that is surely above $d$ (i.e. $b_e \leq d$). \\
Crossing group & One that is crossed by $d$ (i.e. $t_e \leq d < b_e$). \\
\hline
\end{tabular}
\end{table}

\begin{figure}[!t]
	\centering\includegraphics[scale=.4]{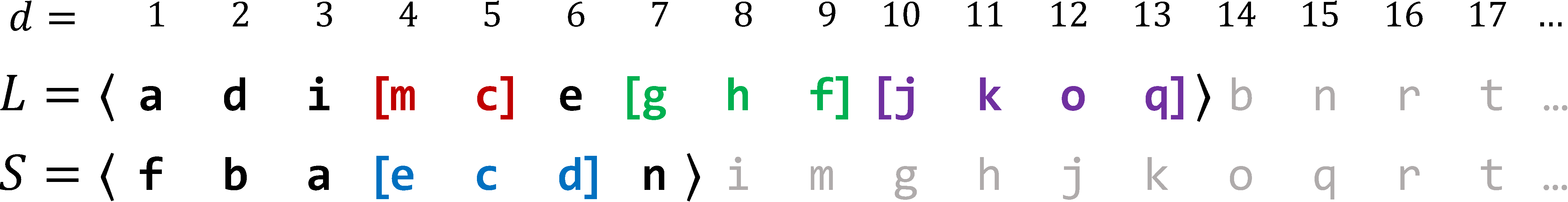}
	\caption{Main example used throughout the paper. Colored letters represent tied items, and square brackets represent tie groups. The grayed-out sections illustrate the arrangement of items that would maximize $\boldsec{RBO_\sub{MAX}}$. Exemplifying notation in Table~\ref{tab:notation}: $\boldsec{l=13, s=7, L_6=\texttt{e}, S^{-1}_\texttt{b}=2, t_{\texttt{g},L}=t_{\texttt{h},L}=t_{\texttt{f},L}=7}$, and $\boldsec{b_{\texttt{e},S}=b_{\texttt{c},S}=b_{\texttt{d},S}=6}$.}\label{fig:example}
\end{figure}

Throughout the paper we will use the notation in Table~\ref{tab:notation} and the example in Figure~\ref{fig:example}. In particular, let $S$ and $L$ be two indefinite rankings of lengths $s$ and $l$, where $S$ is generally shorter than $L$.
\citet{Webber2010} defined their \emph{overlap}, up to a depth $d$, as the number of items that are in common between the two rankings:
\begin{equation}
	X_{S,L,d} = \left|S_{:d} \cap L_{:d}\right|~,\label{eq:overlap}
\end{equation}
where $S_{:d}=\{e~:~\posS \leq d\}$ represents the set of items in $S$ that are ranked at or above the evaluation depth $d$; throughout the paper we refer to these items as \emph{active} in $S$ given $d$.
The proportion of active items that overlap is called the \emph{agreement}:
\begin{equation}
	A_{S,L,d} = \frac{X_{S,L,d}}{d}~.	\label{eq:agreement}
\end{equation}
In the example from Figure~\ref{fig:example}, only item \texttt{a} overlaps at depth 3 and thus $A_3=1/3$. The Rank-Biased Overlap is then defined as the infinite and weighted sum of the agreements at all depths:
\begin{equation}
	RBO_{S,L,p} = \frac{1-p}{p} \sum\nolimits_{d=1}^\infty A_{S,L,d}\cdot p^d~,\label{eq:rbo}
\end{equation}
where $p^d$ is the weight given to the agreement at depth $d$, and the $(1-p)/p$ term ensures that $RBO$ is bounded in the range $[0, 1]$. Fully disjoint rankings result in $RBO=0$, while identical rankings result in $RBO=1$. This is because agreement would be 0 and 1, respectively, at all depths.

The parameter $p$ is called \emph{persistence}, and it determines how steep the decline in weights is: a small $p$ places a very high weight at the top of the ranking compared to the bottom, while a large $p$ flattens the decay so that the weight of deep items is not as small compared to those at the top.
For a full account on the properties of $RBO$, such as metricity, the reader is referred to~\cite{Webber2010}.

\section{Ties in Rank-Biased Overlap}\label{sec:ties}

As discussed in Section~\ref{ssec:ties}, the original view on ties by \citet{Webber2010} is that tied items occur at the same rank, while the view traditionally taken in the Statistics literature is that a tie represents uncertainty as to which item goes first, that is, a loss of information. In the subsequent subsections we first describe \citeauthor{Webber2010}'s approach, which we call $w$-variant, and then fully develop both the $a$- and $b$-variants.

\subsection{$\boldsec{w}$-variant: $\boldsec{RBO^{\!w}}$}\label{ssec:ties-w}

\citet{Webber2010} assumed that all items tied in a group occur at the same rank, namely the top rank of the group. Rankings where tied items are assigned ranks in this way are typically known as \emph{sports} rankings.\footnote{For example, two athletes tied at position 2 share the same rank 2. The next athlete would be at position 4, so that rank 3 is simply unassigned.} In essence, they assumed that items that are tied \emph{really} occur at the same rank.

From this point of view, when considering a group crossed by the evaluation depth $d$, all its items will be active because they are assumed to occur at the top of the group, that is, $\posS=\topS$. Thus, under the sports ranking assumption, $S_{:d}$ may contain more than $d$ items, in particular all items that are either above the depth or tied with it. \citeauthor{Webber2010} then modified the definition of agreement as:
\begin{equation}
	A_{S,L,d}^w = \frac{2\cdot X_{S,L,d}}{|S_{:d}|+|L_{:d}|}~.\label{eq:agreement-w}
\end{equation}
In the example from Figure~\ref{fig:example}, $S_{:5}$ includes item \texttt{d} because it belongs to a crossing group, which happens to increase overlap because it is also found in $L_{:5}$. The denominator is thus 11.

This new definition of agreement may just be plugged into \eqref{eq:rbo} to calculate a tie-aware variant of $RBO$ under the assumption of sports rankings. We will refer to this variant as $RBO^{\!w}$. Note that in the absence of ties $A^{\!w}$ reduces to $A$, so $RBO^{\!w}=RBO$ as expected.

At this point we hint at the gap in the work of \citet{Webber2010}. Indeed, $A^{\!w}$ can be directly used for computing the full $RBO$ score on infinite rankings, but it remains unclear how to use it when computing $RBO$ from prefix evaluation (i.e. how to compute $RBO^{\!w}_\sub{EXT}$ and its bounds). This is because the relevant equations in their work, as well as their rationale, are always expressed in terms of overlap divided by a constant equal to the evaluation depth $d$, not in terms of agreement. The tie-aware $A^{\!w}$ in \eqref{eq:agreement-w} has a different functional form altogether, so one can not simply use their equations to compute $RBO^{\!w}$. This will become more evident next, when we introduce the $a$- and $b$-variants, as well as in Section~\ref{sec:prefix} when we actually deal with prefix evaluation.

\subsection{$\boldsec{a}$-variant: $\boldsec{RBO^a}$}\label{ssec:ties-a}

For the problem of rank correlation, the first way of handling ties under the semantics of uncertainty asked this question: what is the average correlation across all possible permutations of the ties?
This interpretation was followed by \citet{woodbury1940rank} to define a tie-aware variant of Spearman's $\rho$, later by \citet{kendall1945treatment} to define his $\tau_a$, and recently by \citet{urbano2017treatment} to formulate a tie-aware variant of \citeauthor{yilmaz2008new}'s $\tau_{ap}$, namely $\tau_{ap,a}$.

In the same spirit, we ask this question: what is the average agreement among all possible permutations of the ties? This is:
\begin{align}
	A_{S,L,d}^a &=\frac{1}{|\PS||\PL|}\sum_{\permS\in \PS}\sum_{\permL\in \PL}\frac{X_{\tilde{S},\tilde{L},d}}{d}~,\label{eq:agreement-a-1}
\end{align}
where $\tilde{S}$ refers to a single permutation of the tied items in $S$, and $\PS$ refers to all possible such permuted rankings.\footnote{For instance, the possible permutations of ranking $\langle \texttt{A [B C][D E] F}\rangle$ are $\langle \texttt{A B C D E F}\rangle, \langle \texttt{A B C E D F}\rangle, \langle \texttt{A C B D E F}\rangle$ and $\langle \texttt{A C B E D F}\rangle$.} In our example, the number of permutations are 288 and 6 for $L$ and $S$, respectively. Note that the overlap is still defined as in \eqref{eq:overlap}, only that it is computed between permuted rankings instead of the originals.

Computing \eqref{eq:agreement-a-1} by enumerating all possible permutations would be extremely expensive because the number of permutations grows factorially with the number of ties. In order to derive a simple expression that does not require enumeration, we begin by reformulating overlap as follows: instead of calculating the size of the intersection between the items active at depth $d$, we will count the number of items that are active in both rankings:
\begin{align}
	X_{S,L,d} &= \sum_{e\in\Omega}\ind{\{\posS\leq d~ \land ~\posL\leq d\}}\notag \\
	&= \sum_{e\in\Omega} \ind{\{\posS\leq d\}} \cdot \ind{\{\posL\leq d\}}~,\label{eq:overlap-a-1}
\end{align}
where $\Omega\!=\!\{S \cup L\}$ is the set of all items, and $\mathbb{I}$ is the Iverson bracket (i.e. $\ind{P}\!=\!1$ if $P$ is true, 0 otherwise). As expected, an item will contribute to overlap only if it appears at or above $d$ in both rankings. In our example, the summand for item \texttt{b} is $1\cdot 0$ at $d\!=\!3$ (i.e. does not overlap), whereas for item \texttt{a} it is $1\cdot 1$ (i.e. it does overlap).

We can now plug the overlap between two permutations, as defined in \eqref{eq:overlap-a-1}, into the agreement averaged across permutations in~\eqref{eq:agreement-a-1}. After minor rearranging, we obtain:
\begin{equation}
	A_{S,L,d}^a = \frac{1}{d}\sum_{e\in\Omega} {\sum_{\permS\in \PS}\frac{\ind{\{\posSt\leq d\}}}{|\PS|}}{\sum_{\permL \in \PL} \frac{\ind{\{\posLt\leq d\}}}{|\PL|}}~.\label{eq:agreement-a-2}
\end{equation}
Note that the last two summations represent, respectively, the fraction of permutations of $S$ and $L$ where the item $e$ is active. There are three possibilities for an arbitrary item and ranking:
\begin{enumerate}
	\item Inactive: an item or group that is entirely \emph{below} the evaluation depth $d$ (i.e. $d<t_e$) will remain below in all permutations and will never contribute to overlap. In our example, items \texttt{e}, \texttt{c}, \texttt{d} and \texttt{n} are inactive in $S$ at $d=3$.
	\item Active: an item or group that is entirely \emph{at or above} the evaluation depth $d$ (i.e. $b_e\leq d$) will remain above in all permutations and will always be able to contribute to overlap. In our example, items \texttt{a}, \texttt{d}, \texttt{i}, \texttt{m} and \texttt{c} are active in $L$ at $d=5$.
	\item Crossing: an item within a crossing group (i.e. $t_e\!\leq\!d\!<b_e$) will be able to contribute to overlap in as many permutations as it is placed at or above $d$. The item will appear in position $t_e$ a total of $\left(b_e-t_e\right)!$ times, at position $t_e+1$ another $\left(b_e-t_e\right)!$ times, and so on. Only $\left(d-t_e+1\right)$ positions will make the item active, which happens in $\left(d-t_e+1\right)\cdot\left(b_e-t_e\right)!$ permutations. Because there are a total of $\left(b_e-t_e+1\right)!$ permutations of the group, the item's total contribution to overlap is $\left(d-t_e+1\right)/\left(b_e-t_e+1\right)$. In our example, item \texttt{e} in $S$ appears at each of ranks 4, 5 and 6 in 1/3 of the permutations. At depth 5 there are two slots available for the items in the crossing group (i.e. 4 and 5), so each of its items will have a contribution of 2/3.
\end{enumerate}

At a given depth $d$, we can therefore define the total contribution of an item $e$ across permutations as:
\begin{equation}
	c_{e|d} = 
	\begin{cases}
		0 & d < t_e~\text{(inactive)}\\
		1 & b_e\leq d~\text{(active)}\\
		\frac{d-t_e+1}{b_e-t_e+1} & \text{otherwise (crossing)}
	\end{cases}~.\label{eq:cont-a}
\end{equation}
In doing so, we are implicitly assuming that the first unseen item is not tied with the last one seen, as otherwise it should have been seen in the prefix. In the example, \texttt{b} is assumed to not be tied with the last tie group in $L$, so that $c_{L|d}$ can be computed from the prefix only. The contribution of item \texttt{c} in $S$ would be 0 for depths 1 to 3, 1/3 and 2/3 for depths 4 and 5, and 1 for depths 6 and beyond.

Back in \eqref{eq:agreement-a-2}, we can now replace each of the last two summations with the corresponding contributions of $e$ in $S$ and $L$, avoiding the need for enumerating all permutations. All together, a simple and efficient formulation of the agreement for $RBO^a$ is:
\begin{equation}
	A_{S,L,d}^a = \frac{1}{d}\sum_{e\in\Omega} \contS \cdot \contL~.\label{eq:agreement-a}
\end{equation}
In the absence of ties, note that only the first two cases apply in \eqref{eq:cont-a}, which means that $A^a$ reduces to $A$ because \eqref{eq:overlap-a-1} reduces to \eqref{eq:overlap}, so that $RBO^a\!=\!RBO$. Also note that, by construction, $RBO^a$ gives the right answer to the naive approach of computing bare $RBO$ after breaking ties at random, eliminating unnecessary noise.
In addition, note that \eqref{eq:agreement-a} can be computed efficiently because the summation does not need to enumerate all items in $\Omega$, but only those that are active or crossing in both rankings, that is, those where $c_{e,S|d}>0$ and $c_{e,L|d}>0$. 

\subsection{$\boldsec{b}$-variant: $\boldsec{RBO^b}$}\label{ssec:ties-b}

Still under the semantics of uncertainty, the principle for this variant is that it should account for the amount of information actually available to measure overlap. Because a tie represents uncertainty with respect to the actual rank of items, they can not contribute fully to the measured overlap. This inability to fully contribute should be reflected in the normalization term (i.e. the denominator in the agreement function). This leads to the idea of \emph{measurable} overlap in a ranking, or ``untiedness'' as called by \citet{vigna2015weighted}. Whilst the $a$-variant always expects a full measurable overlap of $d$ regardless of the presence of ties, the $b$-variant should not.

This principle to handle ties was first followed by \citet{student1921experimental} to propose a $b$-variant of Spearman's $\rho$, later by \citet{kendall1945treatment} to define his $\tau_b$, and recently by \citet{vigna2015weighted} and \citet{urbano2017treatment}. In order to define a $b$-variant of the agreement for $RBO$, we thus draw inspiration from Kendall's $\tau_b$ which, for two arbitrary (conjoint) rankings $U$ and $V$ is:
\begin{equation}
	\tau_b(U,V) = \frac{\sum_{i<j}\text{sign}(u_j-u_i)\cdot\text{sign}(v_j-v_i)}{
		\sqrt{\sum_{i<j}\text{sign}(u_j-u_i)^2}\sqrt{\sum_{i<j}\text{sign}(v_j-v_i)^2}}~,\label{eq:tau-b}
\end{equation}
where the numerator quantifies \emph{actual}, observed concordance between the rankings, and the denominator quantifies their \emph{measurable} concordance. From this equation we can recognize how an item pair $(i,j)$ affects $\tau_b$ when it is tied: because the tie represents uncertainty, it does not contribute to the numerator in a positive or negative direction. As for the denominator:
\begin{enumerate}
	\item If the item pair is tied in both rankings, it does not contribute to the denominator either. In this case, the item pair is essentially ignored because it does not bear any information about measurable concordance. As such, $\tau_b$ \emph{can still} be 1 if both rankings tie exactly the same items.
	\item If the item pair is tied in only one ranking, it will still contribute to the denominator on behalf of the other ranking. In this case, the item pair is not completely ignored because it still contributes to the measurable concordance. As such, $\tau_b$ \emph{can not} reach 1 any more.	
\end{enumerate}

Applying the same rationale to $RBO$'s agreement, we recognize that tied items should be inactive until the evaluation depth reaches the bottom rank of the group. This is because, until then, their actual ranks are unknown and it should therefore not be possible for them to contribute to overlap at earlier depths. In our example, at depth 5 we can not know which items are actually at ranks 4 and 5 in $S$; it could be any two of \texttt{e}, \texttt{c} and \texttt{d}. Thus, only items \texttt{f}, \texttt{b} and \texttt{a} can contribute to overlap. In contrast, both \texttt{m} and \texttt{c} can contribute in $L$ because their group is entirely active already at depth 5.

Items in a crossing group should therefore not contribute to the numerator. As for the denominator:
\begin{enumerate}
	\item If both rankings have crossing groups at the same ranks, the amount of untiedness is the same in both, say $n$ (in the example, $n=3$ at depth 4). Therefore, the measurable overlap is at most $n$. As such, agreement \emph{can still} reach 1 if both rankings tie exactly the same ranks, regardless of which items they tie.
	\item If the rankings do not have crossing groups at the same ranks, the amount of active items they contribute to the measurable overlap is different (in the example, 6 from $L$ and 7 from $S$ at depth 7). As such, agreement \emph{can not} reach 1 any more.
\end{enumerate}
Dealing with ties in this way, we can recognize an approach somewhat opposite to that of $A^{\!w}$. Indeed, while the $w$-variant assigns to tied items the top rank of their group (i.e. $\posS=\topS$), here we assign them the bottom rank instead (i.e. $\posS=\botS$). By analogy, let us refer to the resulting ranking with the anadrome ``strops'' ranking, to clearly reflect the reverse of ``sports'' ranking.

\begin{figure}[!t]
\centering\includegraphics[scale=.4]{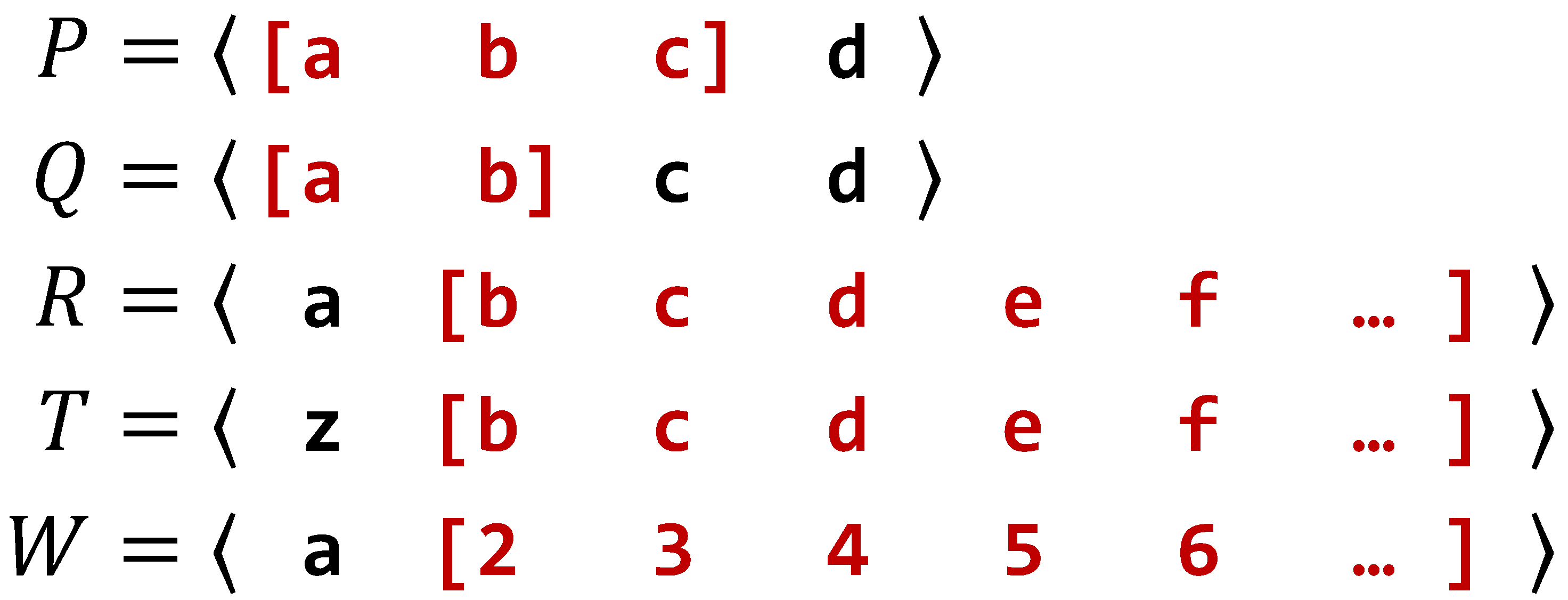}
\caption{Sample rankings that would lead to unexpected $\boldsec{RBO}$ scores with strops rankings. Colored letters represent tied items, and square brackets represent tie groups.}\label{fig:example2}
\end{figure}
Unfortunately, a naive application of the strops ranking would lead to unexpected $RBO$ results, as illustrated in Figure~\ref{fig:example2}:
\begin{itemize}
	\item $R$ vs $W$: the rankings only have the top item in common, so we should intuitively expect an $RBO$ score close to 0. However, agreement is 1 at \emph{every} depth except at the end, where it becomes nearly 0 because all items would contribute to measurable overlap while the actual overlap remains as 1. The final $RBO$ would thus be close to 1 instead of close to 0.
	\item $R$ vs $T$: the top item is different but all the tied items are the same, so we should intuitively expect and $RBO$ score close to 1. However, agreement is 0 at \emph{every} depth except at the end, where it becomes nearly 1. The final $RBO$ would therefore be close to 0 instead of close to 1.
	\item $P$ vs $Q$: the top items are tied in both rankings, which means that at the earliest depths there is no actual overlap, but no measurable overlap either. This would naturally lead to an undefined agreement at those depths, ultimately resulting in an undefined $RBO$. 
\end{itemize}

From these examples, we can make two observations:
\begin{itemize}
	\item O1: we need to ``look inside'' the tie groups to distinguish between simply tying the same items (e.g. $R$ vs $T$) and tying different items altogether (e.g. $R$ vs $W$).
	\item O2: the measurable overlap should always be non-zero to guarantee that agreement is always defined (e.g. $P$ vs $Q$).
\end{itemize}
Note that O1 is also required if we want to ensure that the $RBO$ of a ranking with itself is always 1 regardless of the ties. On the other hand, O2 contrasts with Kendall's $\tau_b$ because it \emph{is} undefined if one ranking is fully tied, reflecting the absence of information to measure concordance in the first place. But for $RBO$ this can not be the case because at least at the very end of the rankings there is no more uncertainty due to ties, so all items are active and contribute to measurable overlap.

In the path towards a solution, we recognize in the actual concordance of $\tau_b$ (i.e. the numerator in \eqref{eq:tau-b}) a similar structure to that of $RBO$'s actual overlap in \eqref{eq:overlap-a-1}: they are both defined as the accumulation of the product of two individual contribution terms, one from each ranking.
To define these individual contribution terms, we note that observation O1 above told us to ``look inside'' the tie groups so that their items have a chance to contribute to actual overlap. This is precisely what we achieved in \eqref{eq:cont-a} when formulating $A^a$, so if we similarly define actual overlap as $\sum_{e\in\Omega} \contS \cdot \contL$, then by analogy to Kendall's $\tau_b$ we can define the measurable overlap of a single ranking as $\sqrt{\sum_{e\in\Omega}c^2_{e|d}}$. Finally, we note that $c_{e|d}$ is always non-zero for every item in $\Omega$, which ensures a non-zero measurable overlap as required by observation O2 above.

All this considered, we propose the following formulation of agreement for $RBO^b$:
\begin{equation}
	A_{S,L,d}^b = \frac{\sum_{e\in\Omega} c_{e,S|d} \cdot c_{e,L|d}}{\sqrt{\sum_{e\in\Omega} c_{e,S|d}^2}\sqrt{\sum_{e\in\Omega} c_{e,L|d}^2}}~,\label{eq:agreement-b}
\end{equation}
which is bounded between 0 and 1 due to the Cauchy–Schwarz inequality. In the absence of ties, note again that only the first two cases apply in \eqref{eq:cont-a} and the denominator equals $d$, which means that $A^b$ reduces to $A$ and therefore $RBO^b=RBO$.

Note that $A^b$ and $A^a$ measure actual overlap at the numerator in the same way, but differ in the measurable overlap at the denominator. In our main example, $A^a_4$ has a measurable overlap of 4, while for $A^b_4$ ranking $S$ contributes $\sqrt{3+1/3}$ and ranking $L$ contributes $\sqrt{3+1/2}$.
As a side product then, $A^a$ is always less than or equal to $A^b$. This relation between the $a$- and $b$-variants is a result of how \citet{kendall1945treatment} connected the work of \citet{student1921experimental} and \citet{woodbury1940rank} for his definitions of $\tau_a$ and $\tau_b$. The numerator is the same, nicely representing the expected amount of concordance across all permutations of the tied items, but the denominators then differ in whether they correct for untiedness or not. This is exactly the relation we have between $A^a$ and $A^b$, and by extension between $RBO^a$ and $RBO^b$.

\section{Prefix Evaluation With Ties}\label{sec:prefix}

As presented in \eqref{eq:rbo}, $RBO$ is defined on infinite rankings, but they are usually truncated, as mentioned in Section~\ref{sec:intro}. This means that rankings actually consist of a seen part or prefix, and an unseen part that extends up to infinity. Therefore, $RBO$ scores have to be computed from a prefix only, ideally accompanied by some quantification of the uncertainty due to the unseen items.
\citet{Webber2010} presented the rationale and equations to compute upper and lower bounds on $RBO$. For the lower bound $RBO_\sub{MIN}$, it is assumed that all items in the unseen parts are disjoint, thus minimizing the agreement. For the upper bound $RBO_\sub{MAX}$, it is assumed that every item in the unseen part of one ranking matches an item in the other one, thus maximizing the agreement. The difference, $RBO_\sub{RES}=RBO_\sub{MAX}-RBO_\sub{MIN}$, directly quantifies the magnitude of the residual. Lastly, they also introduced a point estimate named $RBO_\sub{EXT}$, calculated by extrapolating the agreement measured in the prefixes, assuming the same agreement would be observed throughout the unseen parts.

However, as mentioned already in Section~\ref{ssec:ties-w}, the relevant equations they present (i.e. (11), (30) and (32)), as well as the rationale behind them, are always expressed in terms of overlap, not in terms of agreement. It is clear already with their $A^{\!w}$ in \eqref{eq:agreement-w} that a definition for agreement may have, not only a custom denominator other than simply $d$, but a different functional form altogether. This is now even more evident from $A^a$ in \eqref{eq:agreement-a} and $A^b$ in \eqref{eq:agreement-b}. As a consequence, their equations can \emph{not} be used for prefix evaluation of $RBO$ in the presence of ties and, as will be shown in Section~\ref{ssec:summation-2}, the rationales behind $RBO_\sub{MAX}$ and $RBO_\sub{EXT}$ are actually a bit more involved than it seemed with bare $RBO$ and no ties.

In order to fill this gap and derive a general formulation for $RBO$ and prefix evaluation, we first rewrite the full $RBO$ from \eqref{eq:rbo} by explicitly separating three sections: 1) from depth 1 up to $s$, where both rankings are seen, 2) from $s+1$ up to $l$, where only $L$ is seen, and 3) from $l+1$ up to infinity, where both rankings are unseen:
\begin{equation}
RBO_{S,L,p} = \frac{1-p}{p}\Biggl( \underbrace{\sum_{d=1}^s A_d p^d}_{1} + \underbrace{\sum_{d=s+1}^l A_d p^d}_{2} + \underbrace{\sum_{d=l+1}^\infty A_d p^d}_{3}\Biggl)~.\label{eq:rbo-in-3}
\end{equation}

For simplicity, in the remainder of the paper we will use exclusively the formulations that compute overlap based on the products of individual contributions, such as in~\eqref{eq:overlap-a-1} and \eqref{eq:agreement-a}. Note that the $w$-variant can be easily expressed in this way, for example as:
\begin{equation}
	A_{S,L,d}^w = \frac{2\cdot\sum_{e\in\Omega} \contS \cdot \contL}{
		\sum_{e\in\Omega}\contS + \sum_{e\in\Omega}\contL}~,\label{eq:agreement-w-cont}
\end{equation}
where $\contS=\ind{t_{e,S} \leq d}$, and likewise for $\contL$. 

Note that agreement can be readily measured in the first section because both rankings are seen, but for the second section we need to make an assumption about the unseen items in $S$ and their overlap with $L$. Likewise, in the third section we need an assumption about unseen items in both $S$ and $L$. What assumptions are made, depends on whether we compute $RBO_\sub{MIN}$, $RBO_\sub{MAX}$ or $RBO_\sub{EXT}$.

In addition, we assume there are no ties in the unseen parts. This is necessary for the $w$-variant because prior information about conjointness would otherwise be needed to compute bounds. Indeed, tying all the unseen items when rankings are mostly conjoint would maximize $RBO^{\!w}$ because they would contribute to both the numerator and denominator in \eqref{eq:agreement-w} at earlier depths. On the other hand, tying all unseen items in a mostly non-conjoint case would actually minimize $RBO^{\!w}$ because they would only contribute to the denominator. For the $a$- and $b$-variants, it just makes sense to assume no ties, as they assume the existence of fully untied rankings in the first place; recall that in these variants a tie just reflects an inability to distinguish items that are too close together.

\subsection{Second Section: from $\boldsec{s+1}$ to $\boldsec{l}$}\label{ssec:summation-2}

The agreement in this second section depends on how the unseen items in $S$ are assumed to overlap with $L$, and how the agreement function combines this unseen overlap with both the seen and measurable overlaps, which ultimately depends on the tie-variant.

Let us define this assumed overlap $\tilde{X}_d$ by separating two components: one measuring the actual overlap among the seen items, and another one incorporating the assumed contribution of the unseen items in $S$:
\begin{equation}
	\tilde{X}_d = \underbrace{\sum\nolimits_{e\in \Omega}\contS\cdot\contL}_{\text{seen}} +\underbrace{\sum\nolimits_{k=s+1}^d \tilde{c}_{k,S|d}\cdot\tilde{c}_{k,L|d}}_{\text{unseen}}~.\label{eq:overlap-assumed}
\end{equation}
Note that the first summation is simply the regular overlap, and that an item that only appears in $L$ will \emph{not} contribute here because its $c_{S|d}$ is 0. These unmatched items from $L$ are the ones that have a chance to overlap with the $d-s$ unseen items in $S$ through the second summation. For each of those unseen items in $S$ we assume a contribution $\tilde{c}_{k,S|d}$, and a corresponding non-constant contribution $\tilde{c}_{k,L|d}$ that depends on what item is actually matched in $L$.
This is the point where the derivations by \citet{Webber2010} are not sufficiently general to accommodate ties, because when an item in $L$ is matched in the unseen part of $S$, they give it a unitary contribution to overlap, which is not necessarily correct in the presence of ties and fractional contributions.

In order to compute agreement in each of the three variants, we need to combine the assumed overlap in \eqref{eq:overlap-assumed} with the measurable overlap. In this respect, we note that the measurable overlap contributed by $S$ is always equal to $d$ because unseen items are assumed to be untied. The second summation in \eqref{eq:rbo-in-3} becomes:
\begin{align}
	\left(\sum_{d=s+1}^l \!A_d p^d\right)^w &=
	\sum_{d=s+1}^l \frac{2\cdot \tilde{X}_d}{d+|L_{:d}|} p^d~,\\
	\left(\sum_{d=s+1}^l \!A_d p^d\right)^a &=
	\sum_{d=s+1}^l \frac{\tilde{X}_d}{d} p^d~,~\text{and}\\
	\left(\sum_{d=s+1}^l \!A_d p^d\right)^b &=
	\sum_{d=s+1}^l \frac{\tilde{X}_d}{\sqrt{d}\sqrt{\sum_{e\in\Omega}\contL^2}} p^d~.\label{eq:sum2-ext}
\end{align}

\subsubsection{$RBO_\sub{MIN}$} For the lower bound it is assumed that all unseen items are disjoint. This means that their individual contributions are 0, so they do not contribute to the unseen overlap in any way:\footnote{Setting $\tilde{c}_{k,L|d}=0$ is arbitrary, but this is irrelevant because $\tilde{c}_{k,S|d}$ \emph{must} be 0 anyway.}
\begin{equation}
	\big(\tilde{c}_{k,S|d}\big)_\sub{MIN}=0~,\quad 
	\big(\tilde{c}_{k,L|d}\big)_\sub{MIN} = 0~.
\end{equation}

\subsubsection{$RBO_\sub{MAX}$} 
Every unseen item in $S$ has a unitary contribution because it is untied and it matches an item in $L$. However, the corresponding contribution in $L$ must take into account the order of the unmatched items. Indeed, $RBO$ is maximized when the $k$-th unseen item at rank $s+k$ matches the $k$-th still unmatched item in $L$. These are the grayed-out items in Figure~\ref{fig:example}. For instance, \texttt{m} is the item that maximizes agreement at depth 9, with a contribution $\tilde{c}_{L|d}=1$. Note that the item maximizing agreement at depth 12 can be any of \texttt{j}, \texttt{k}, \texttt{o} and \texttt{q}, for they are in a crossing group at that depth. In the $w$-variant they would have an individual contribution $\tilde{c}_{L|d}=1$, or $\tilde{c}_{L|d}=3/4$ in the $a$- and $b$-variants.

We therefore need to know the sequence of items unique to $L$ that are potentially active, and arranged in the same order; let us refer to these  as $U_d=\langle u_i ~:~c_{u_i,L|d}>0 \wedge c_{u_i,S|d}=0 \wedge c_{u_i,L|d}\geq c_{u_{i+1},L|d}\rangle$. The individual contributions to the unseen overlap are therefore:
\begin{equation}
	\big(\tilde{c}_{k,S|d}\big)_\sub{MAX}=1~,\quad 
	\big(\tilde{c}_{k,L|d}\big)_\sub{MAX} = c_{u_k,L|d}~.
\end{equation}
In the absence of ties, note that all $\tilde{c}_{k,L|d}$ are equal to 1.

\subsubsection{$RBO_\sub{EXT}$} In this case, no specific items are assumed for the unseen ranks of $S$. Instead, the assumption is about their individual contributions to unseen overlap. \citet{Webber2010} decided to set these contributions equal to $A_s$, albeit the corresponding contribution from ranking $L$ was still assumed to be 1. However, and similarly to the case of $RBO_\sub{MAX}$, these contributions are not necessarily 1 when dealing with ties and crossing groups. We take a slightly different approach and assume that an unseen item in $S$ may match, with equal probability, any of the potentially active but still unmatched items in $L$, that is, any of the items in $U_d$. We then ask for the expected value of the joint contribution at rank $k$:
\begin{align}
	\mathrm{E}\!\left[c_{S|d}\cdot c_{L|d}~\big|~k\right] &= P(\mathrm{unmatch}~|~k)\cdot 0~+ \notag\\
	& + P(\mathrm{match}~u_1~|~k)\cdot 1\cdot c_{u_1,L|d}~+ \notag\\
	& + P(\mathrm{match}~u_2~|~k)\cdot 1\cdot c_{u_2,L|d}~+ \dots = \notag\\
	& = P(\mathrm{match}~|~k)\cdot \mathrm{E}\!\left[c_{L|d}~\big|~U_d\right]~, \label{eq:expected-contribution}
\end{align}

We note here that $P(\mathrm{match}~|~k)$ is precisely where extrapolation happens via $A_s$; indeed, agreement can be interpreted as the probability that an item chosen at random appears in both rankings. From this view, $A_s$ is therefore not the assumed contribution of the unseen item in $S$, which is always 1 because we assume it to be untied, but rather the probability that it matches something in $L$. Nonetheless, the individual contributions may be defined as follows to incorporate in \eqref{eq:overlap-assumed}:
\begin{equation}
	\big(\tilde{c}_{k,S|d}\big)_\sub{EXT}=A^*_s~,\quad 
	\big(\tilde{c}_{k,L|d}\big)_\sub{EXT} = \frac{1}{|U_d|}\sum\nolimits_{e\in U_d} c_{e,L|d}~.
\end{equation}
Note that the agreement at $s$ depends on what tie-aware variant is being used, which we indicate with the star $*$. In the absence of ties, all contributions from $L$ are also unitary, which means that all $\tilde{c}_{L|d}$ are ultimately equal to 1, making the total contribution of an unseen item equal to $A_s$, as formulated for bare $RBO$.

Finally, we must note that the agreement in \eqref{eq:sum2-ext} is still bounded by 1. Loosening notation, $\sum{\tilde{c}_{S|d}\tilde{c}_{L|d}}$ is bounded by $\sqrt{\smash[b]{\sum\tilde{c}_{S|d}^2}}\sqrt{\smash[b]{\sum\tilde{c}_{L|d}^2}}$ due to the Cauchy–Schwarz inequality. The first term is itself bounded by $\sqrt{k}$ because $A_s^*\leq 1$, and the second term is bounded by $\sqrt{\smash[b]{\sum c_{L|d}^2}}$ due to Jensen's inequality.

\subsection{Third Section: from $\boldsec{l+1}$ to $\boldsec{\infty}$}\label{ssec:summation-3}

Regarding the third summation in \eqref{eq:rbo-in-3}, we first note that all the items seen in the prefixes are active at depths $l$ and beyond, so the seen overlap at those depths is independent of the tie variant.

\subsubsection{$RBO_\sub{MIN}$} All unseen items in the third section are assumed to be disjoint, so the overlap will remain constant and equal to $X_l$:
\begin{equation}
	\left(\sum_{d=l+1}^\infty \!A_d p^d\right)_\sub{MIN} \!=\! \sum_{d=l+1}^\infty \frac{X_l}{d} p^d = X_l \left[ \ln \left(\frac{1}{1\!-\!p}\right) - \sum_{d=1}^l \frac{p^d}{d} \right]~.
\end{equation}

\subsubsection{$RBO_\sub{MAX}$} The $l-s$ unseen items in $S$ from the second section are assumed to match an item in $L$, so the assumed overlap at depth $l$ becomes $X_l+l-s$. After $l$, every unseen item in $L$ is assumed to match an unmatched item in $S$ and vice-versa, thus contributing $+2$ to the overlap (in our example, this happens at depths 14 and 15). This continues until all the remaining unmatched items are placed after the prefixes, which happens at depth $f=l+s-X_l$ (in our example, $f=15$). After $f$, it is assumed that the same item would appear in both rankings, thus continuing the full agreement indefinitely. The third summation for $RBO_\sub{MAX}$ is therefore split in two subsections: from $l+1$ up to $f$, where overlap increases by 2 at each step, and from $f+1$ up to $\infty$, where full agreement is assumed:
\begin{align}
	\left(\sum_{d=l+1}^\infty \!A_d p^d\right)_\sub{MAX} &\!=\! \sum_{d=l+1}^f \frac{2d-l-s+X_l}{d}p^d + \sum_{d=f+1}^\infty p^d \nonumber \\
	&\!=\! \sum_{d=l+1}^f \frac{2d-l-s+X_l}{d}p^d + \frac{p^{f+1}}{1-p}~.
\end{align}

\subsubsection{$RBO_\sub{EXT}$} Recall that we have to extrapolate the agreement at $l$ to all subsequent depths up to infinity. Because all items are finally active in this third section, we have that $\tilde{c}_{L|d}$ is always 1. This means that the assumed overlap $\tilde{X}_l$ that we extrapolate equals $X_l + A_s^*(l-s)$. In addition, both $S$ and $L$ have the same full contribution to measurable overlap at the denominators, which means that the extrapolated agreement at $l$ takes the same form for all tie-variants: $\left(X_l + A_s^*(l-s)\right)\!/l$. Extrapolating this agreement up to infinity, we have the third summation for $RBO_\sub{EXT}$:

\begin{align}
	\left(\sum_{d=l+1}^\infty A_d p^d\right)^*_\sub{EXT} &\!=\! \sum_{d=l+1}^\infty \frac{X_l + A^*_s(l-s)}{l} p^d \nonumber\\
	&\!=\! \frac{X_l + A^*_s(l-s)}{l} \cdot \frac{p^{l+1}}{1-p}~.
\end{align}
In the absence of ties, all these formulations reduce to bare $RBO$.

\section{Experimental Demonstrations}\label{sec:exp}

We now illustrate the use of $RBO$ in the presence of ties, emphasizing the differences between each of the three tie-aware variants and bare $RBO$, thus showing the importance of handling ties explicitly. Because using $RBO$ normally involves the calculation of the extrapolated score, we focus only on $RBO_\sub{EXT}$. Revisiting Section~\ref{sec:prefix}, we see that all variants are essentially the same with respect to the bounds, where the only difference is the definition of $\big(\tilde{c}_{k,L|d}\big)_\sub{MAX}$. These differences are negligible and make $RBO_\sub{RES}$ depend mostly on the evaluation depth, not on the handling of ties.

%
%
%

\subsection{TREC Data}\label{ssec:trec}
\def\figscale{.48}
\begin{figure}[!t]\small
	\centering
	a) Ties broken at random in $RBO$\\
	\includegraphics[scale=\figscale]{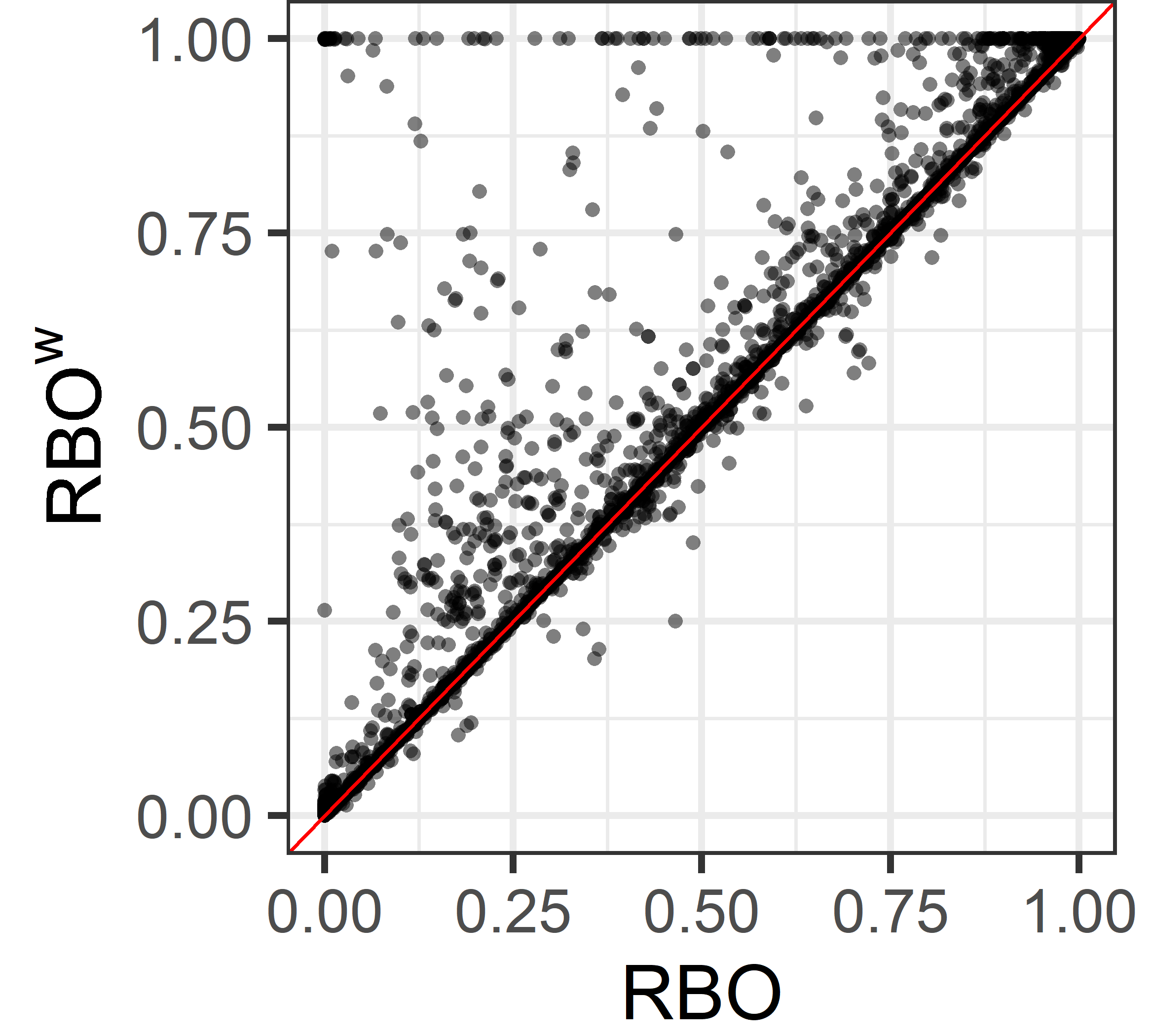}%
	\includegraphics[scale=\figscale]{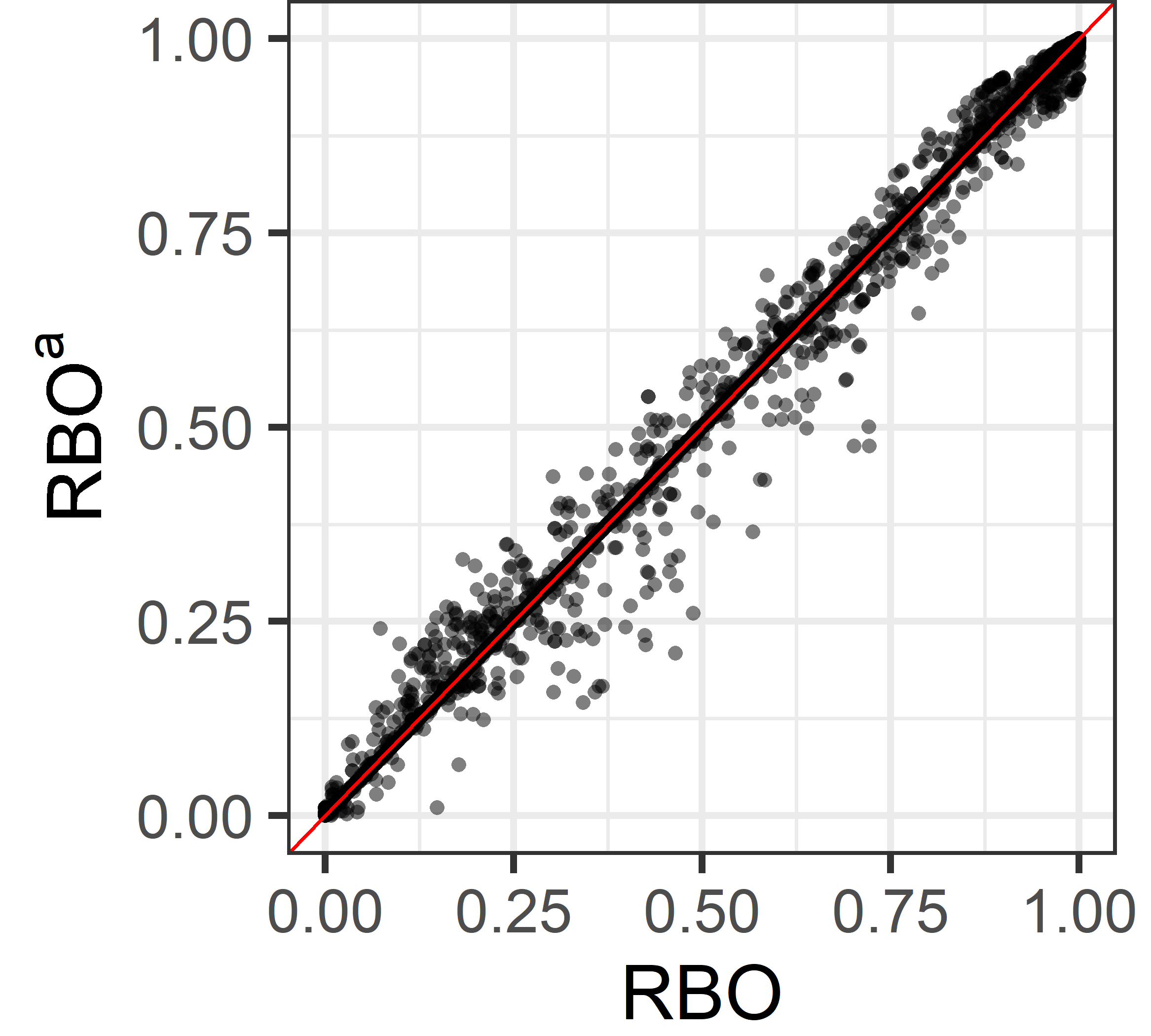}%
	\includegraphics[scale=\figscale]{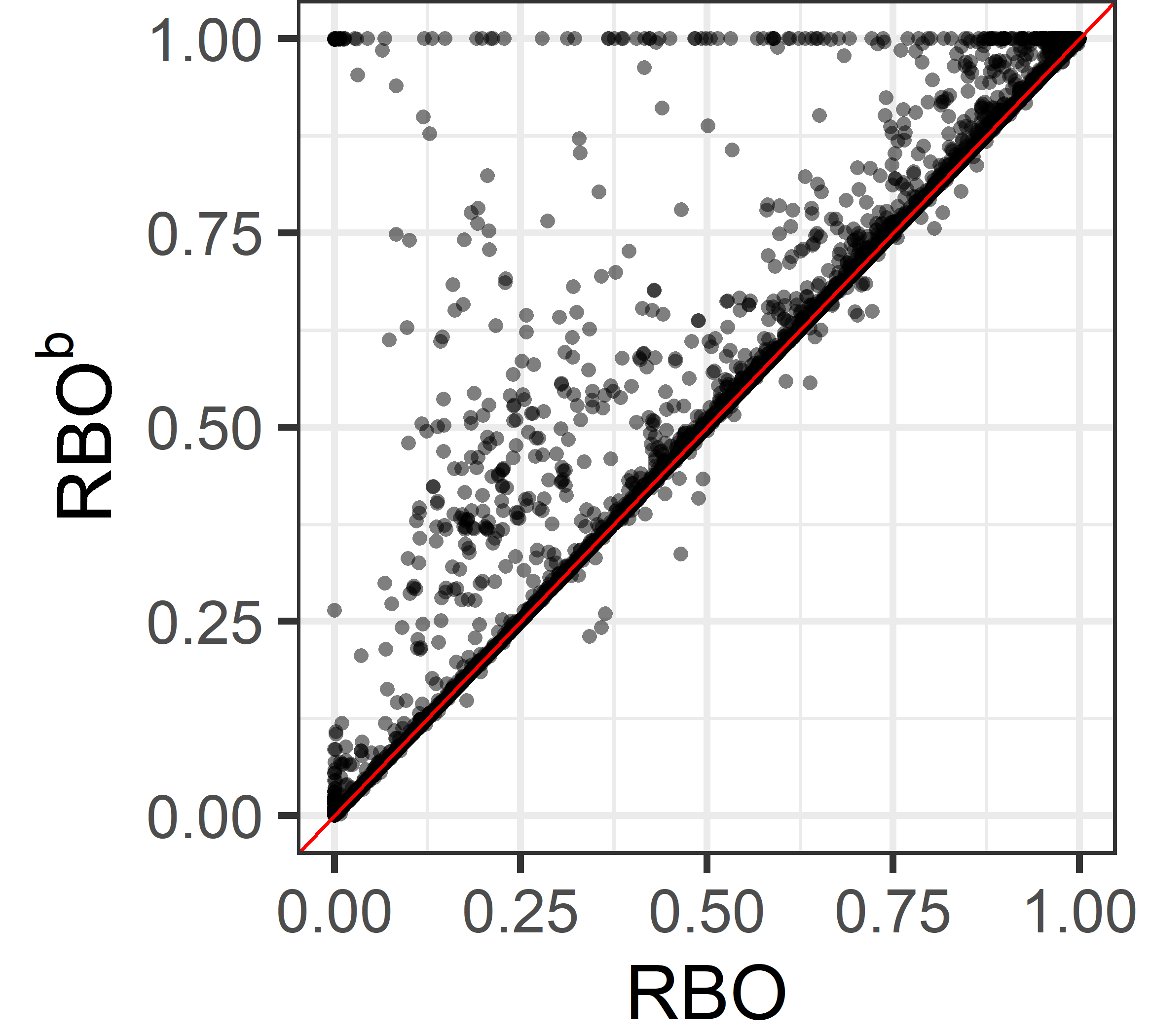}\\%
	b) Ties broken by doc ID in $RBO$\\
	\includegraphics[scale=\figscale]{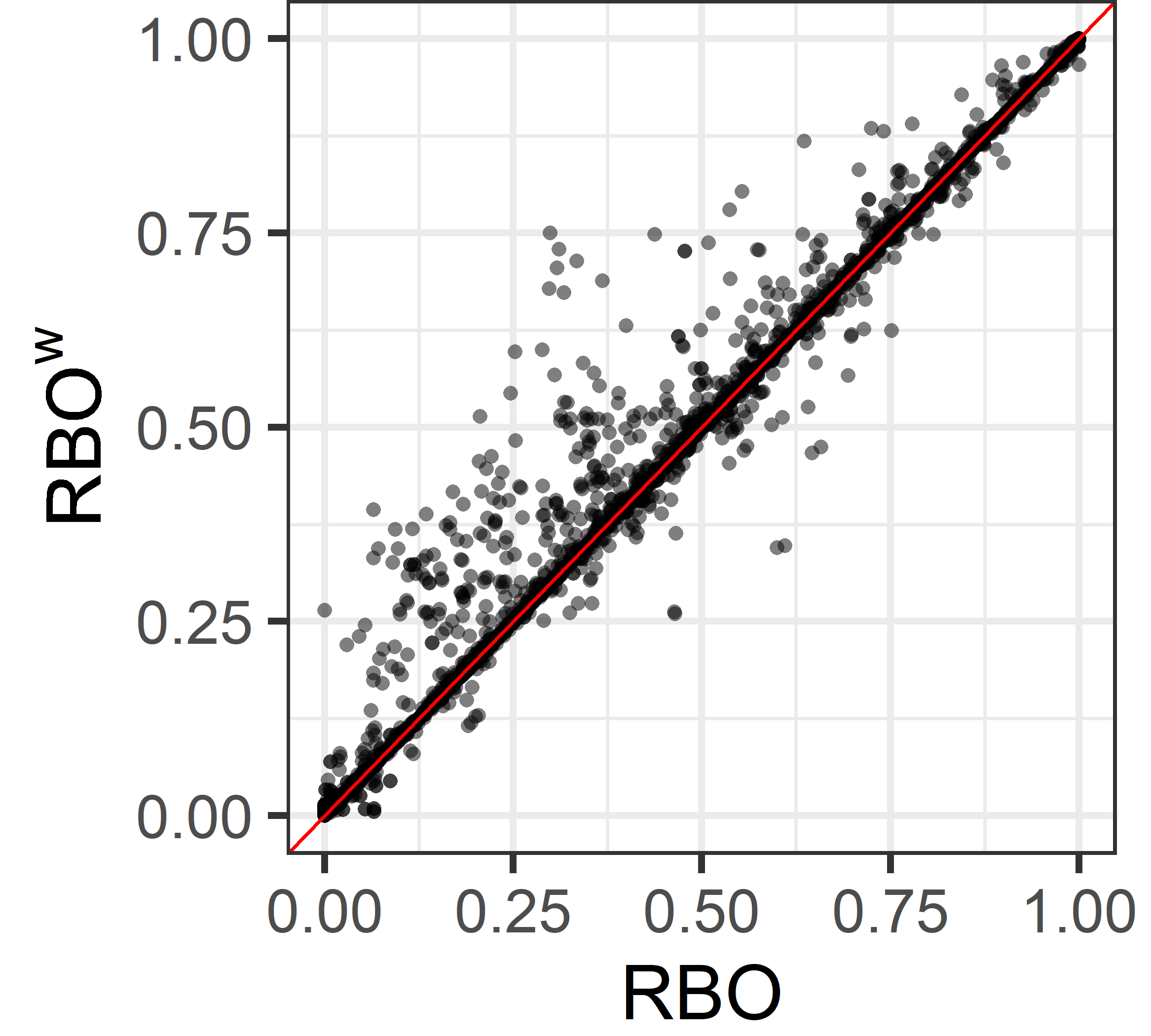}%
	\includegraphics[scale=\figscale]{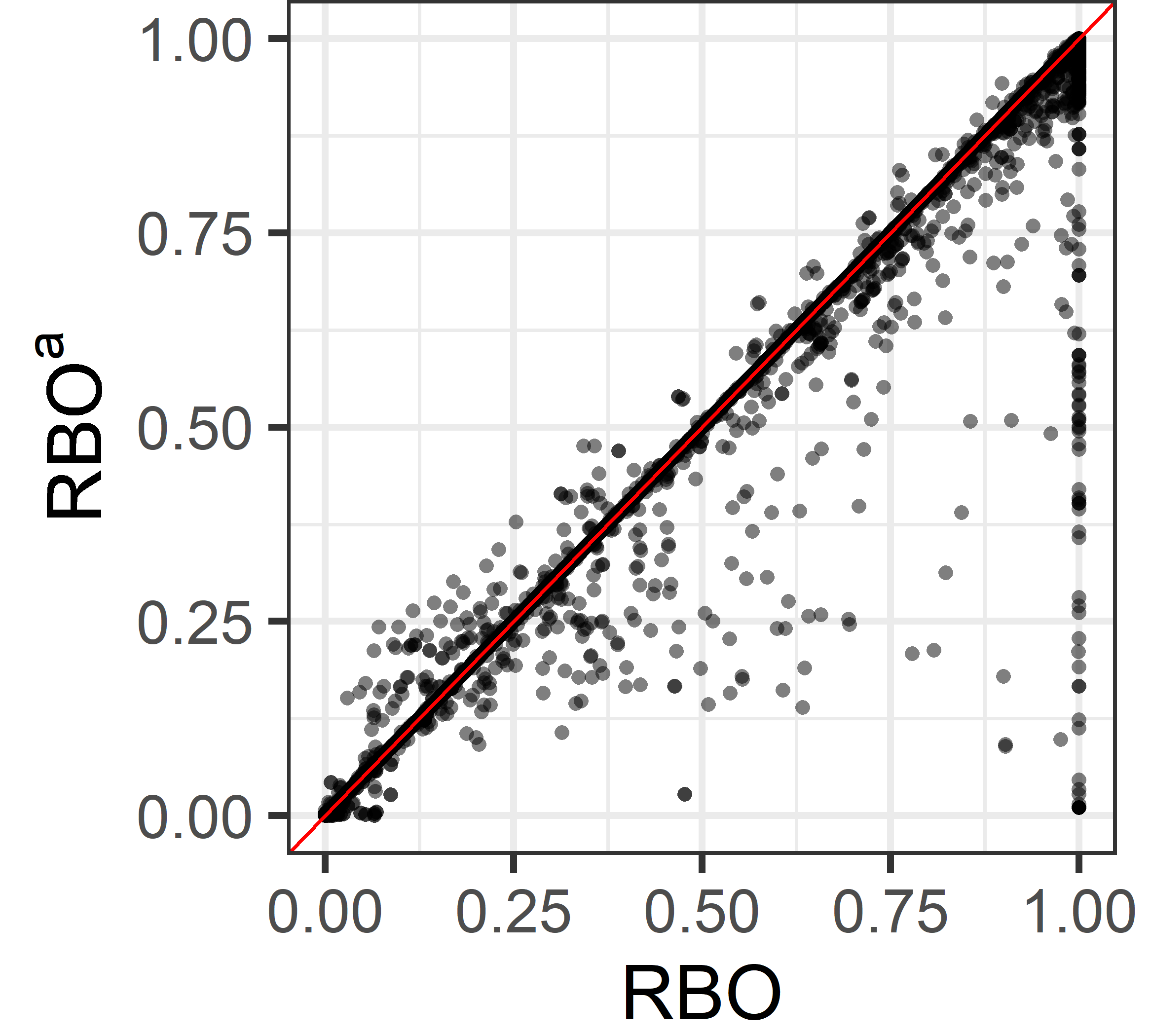}%
	\includegraphics[scale=\figscale]{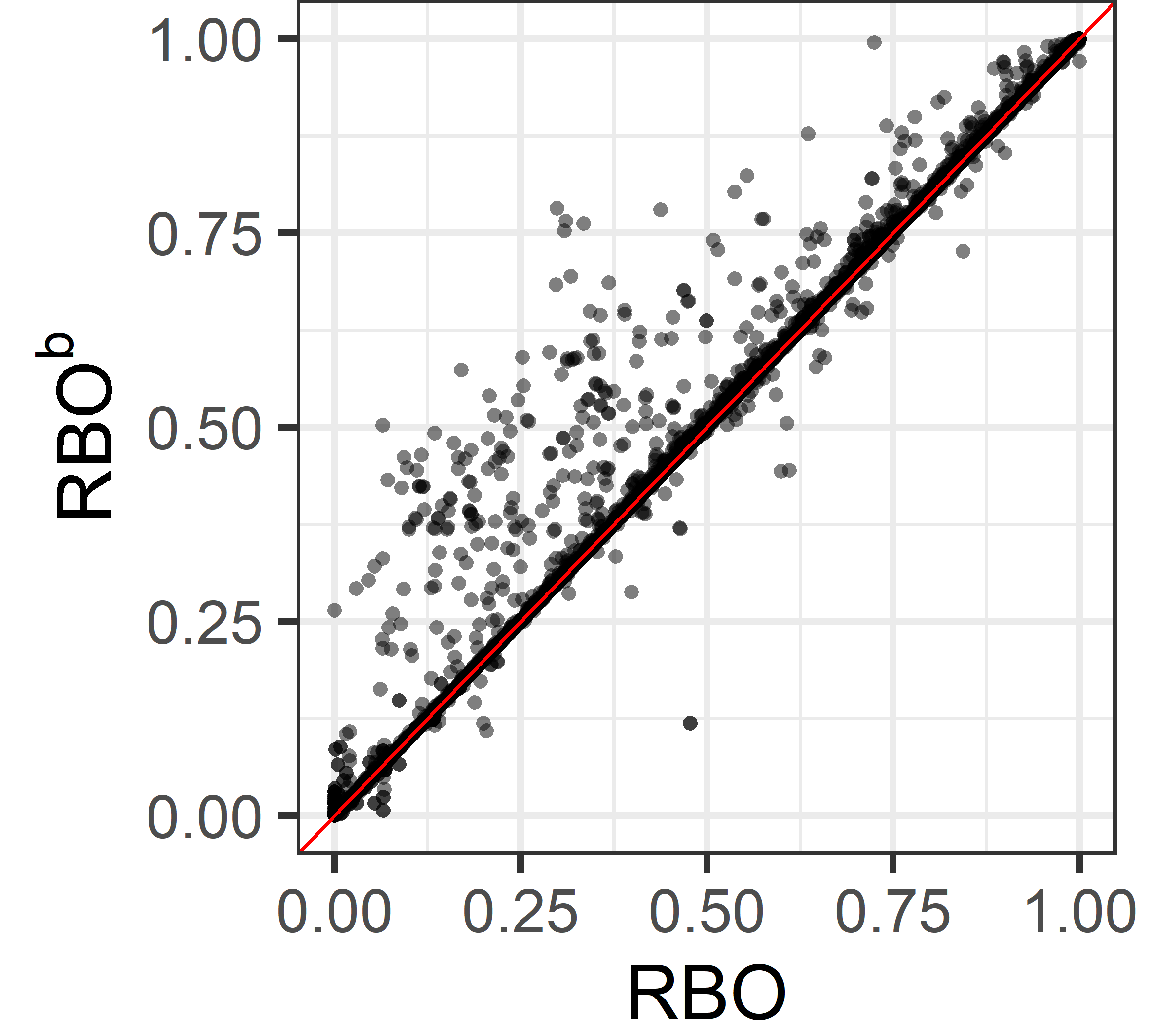}
	\caption{Differences between bare $\boldsec{RBO}$ and each of the three tie-aware variants. TREC data, $\boldsec{p\!=\!0.9}$.}\label{fig:trec}
\end{figure}

A common use case for $RBO$ is comparing the document rankings returned by an experimental system with those of a baseline. From all adhoc runs in the TREC 2009--2014 Web track (see Table~\ref{tab:trec-summary}), we compared the rankings between all 255 pairs of runs by the same group and over the 50 topics, for a grand total of 12,750 pairs of rankings. Of those, 9,256 contained ties, which are the ones we report about. The average ranking length was 978 documents, with a maximum of 1,000. For every pair of rankings, we computed the tie-unaware $RBO$, breaking ties at random and by doc ID, as well as $RBO\!^w$, $RBO^a$ and $RBO^b$. The chosen values for $p$ were 0.8, 0.9 and 0.95, thus setting the expected number of results compared by the $p$-persistent user to 5, 10 and 20 documents, respectively.

Figure~\ref{fig:trec} compares all $RBO$ scores for $p=0.9$. All variants are of course highly correlated, but we can observe some striking differences. Focusing first on $RBO^{\!w}$ and $RBO^b$, we see that they are closer to bare $RBO$ when breaking ties by doc ID. As mentioned earlier, this is expected because both rankings will have these tied items artificially sorted in the same order when presented to bare $RBO$, inflating the result. On the other hand, $RBO^{\!w}$ and $RBO^b$ are specifically designed to deal with these items, and if they happen to be similarly distributed in both rankings they will actually contribute positively.
In contrast, $RBO^a$ is closer to bare $RBO$ when breaking ties at random. Again, this is expected, as they are by construction equal on expectation (see Section~\ref{ssec:ties-a}). As a consequence, $RBO^a$ is generally lower than $RBO$ breaking ties by doc ID because these documents contribute in different directions across permutations.
The key takeaway is that there can be \emph{very} large differences among variants, so a sensible decision should be made as to which one should be computed depending on the specific meaning of ties.

Table~\ref{tab:trec} provides a summary of these differences among variants. In particular, we can see that, while differences are small on average, there are \emph{very} large cases. Overall, the table confirms that $RBO^{\!w}$ and $RBO^b$ are most different from bare $RBO$ when breaking ties at random, whilst that makes it closer to $RBO^a$. Recall here that the table reports \emph{absolute} differences; signed differences with $RBO^a$ are actually 0, as expected. Another way to look at deviations is by classifying them in large (more than 0.1), medium (between 0.01 and 0.1) or small (less than 0.01), which roughly translates into differences in the first, second, or third decimal digit of a reported $RBO$ score, respectively; the first two are identified as M and L in Table~\ref{tab:trec}.
As can be seen, about 4\% of the observed differences are large, while about 8\% are of medium size. This means that the strategy followed to deal with ties brings a substantial difference in about 12\% of the comparisons between rankings.

\begin{table}[!t]
	\caption{Summary of differences between bare $\boldsec{RBO}$ and each of the three tie-aware variants. M for medium differences in $\boldsec{(0.01,0.1]}$, and L for large in $\boldsec{(0.1,1]}$. TREC data.}\label{tab:trec}
	\centering\small\setlength\tabcolsep{1mm}\begin{tabular}{|r|rrrr|rrrr|rrrr|}
		\multicolumn{13}{c}{a) Ties broken at random in $RBO$}\\\hline
		& \multicolumn{4}{c|}{$\left|RBO-RBO\!^w\right|$} & \multicolumn{4}{c|}{$\left|RBO-RBO^a\right|$} & \multicolumn{4}{c|}{$\left|RBO-RBO^b\right|$} \\ 
		$p$& Avg. & Max. & M & L & Avg. & Max. & M & L & Avg. & Max. & M & L \\
		\hline
		0.8 & 0.02 & 1 & 8\% & 5\% & 0.01 & 0.43 & 7\% & 2\% & 0.02 & 1 & 7\% & 5\% \\
		0.9 & 0.02 & 1 & 9\% & 4\% & $<$.01 & 0.26 & 7\% & 1\% & 0.02 & 1 & 8\% & 4\% \\
		0.95 & 0.01 & 1 & 9\% & 3\% & $<$.01 & 0.15 & 6\% & $<$1\% & 0.01 & 1 & 8\% & 3\% \\
		\hline
		\multicolumn{13}{c}{}\\
		\multicolumn{13}{c}{b) Ties broken by doc ID in $RBO$}\\\hline
		& \multicolumn{4}{c|}{$\left|RBO-RBO\!^w\right|$} & \multicolumn{4}{c|}{$\left|RBO-RBO^a\right|$} & \multicolumn{4}{c|}{$\left|RBO-RBO^b\right|$} \\ 
		$p$ & Avg. & Max. & M & L & Avg. & Max. & M & L & Avg. & Max. & M & L \\
		\hline
		0.8 & 0.01 & 0.60 & 6\% & 2\% & 0.01 & 1 & 6\% & 4\% & 0.01 & 0.63 & 5\% & 3\% \\
		0.9 & 0.01 & 0.45 & 6\% & 2\% & 0.01 & 0.99 & 8\% & 2\% & 0.01 & 0.48 & 5\% & 2\% \\
		0.95 & $<$.01 & 0.31 & 5\% & 1\% & 0.01 & 0.98 & 8\% & 2\% & 0.01 & 0.45 & 5\% & 2\% \\
		\hline
	\end{tabular}
\end{table}

\begin{figure}[!t]
	\includegraphics[scale=\figscale]{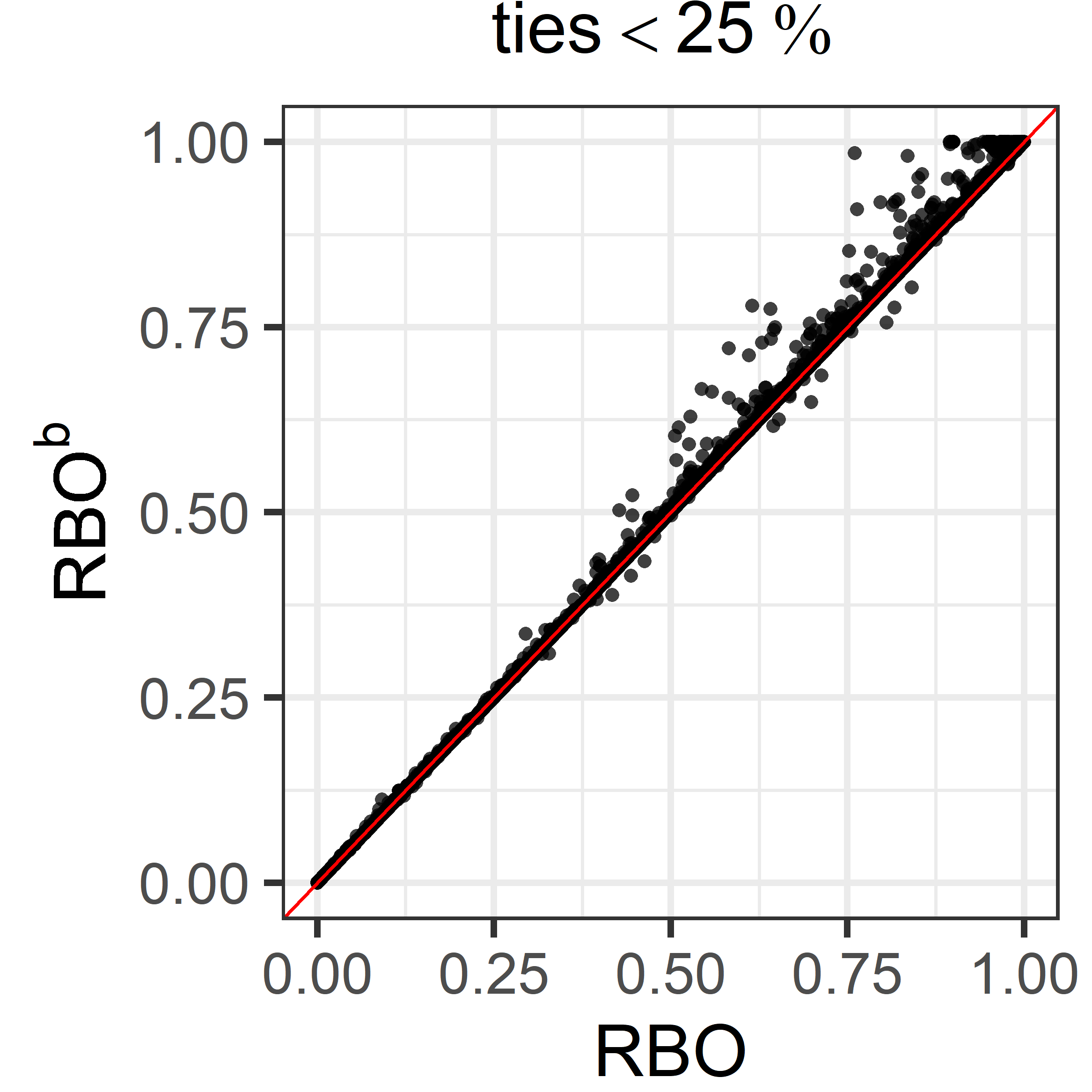}%
	\includegraphics[scale=\figscale]{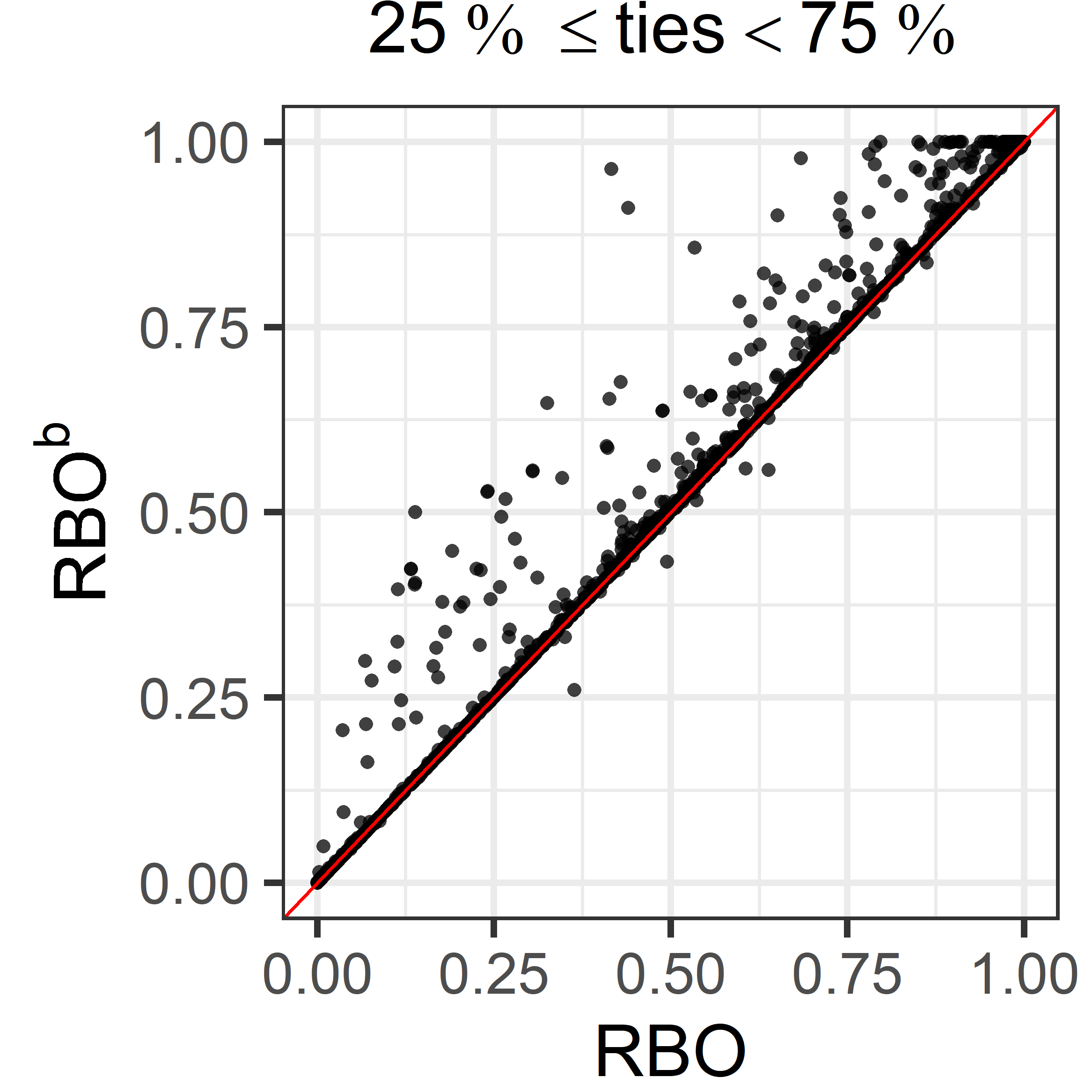}%
	\includegraphics[scale=\figscale]{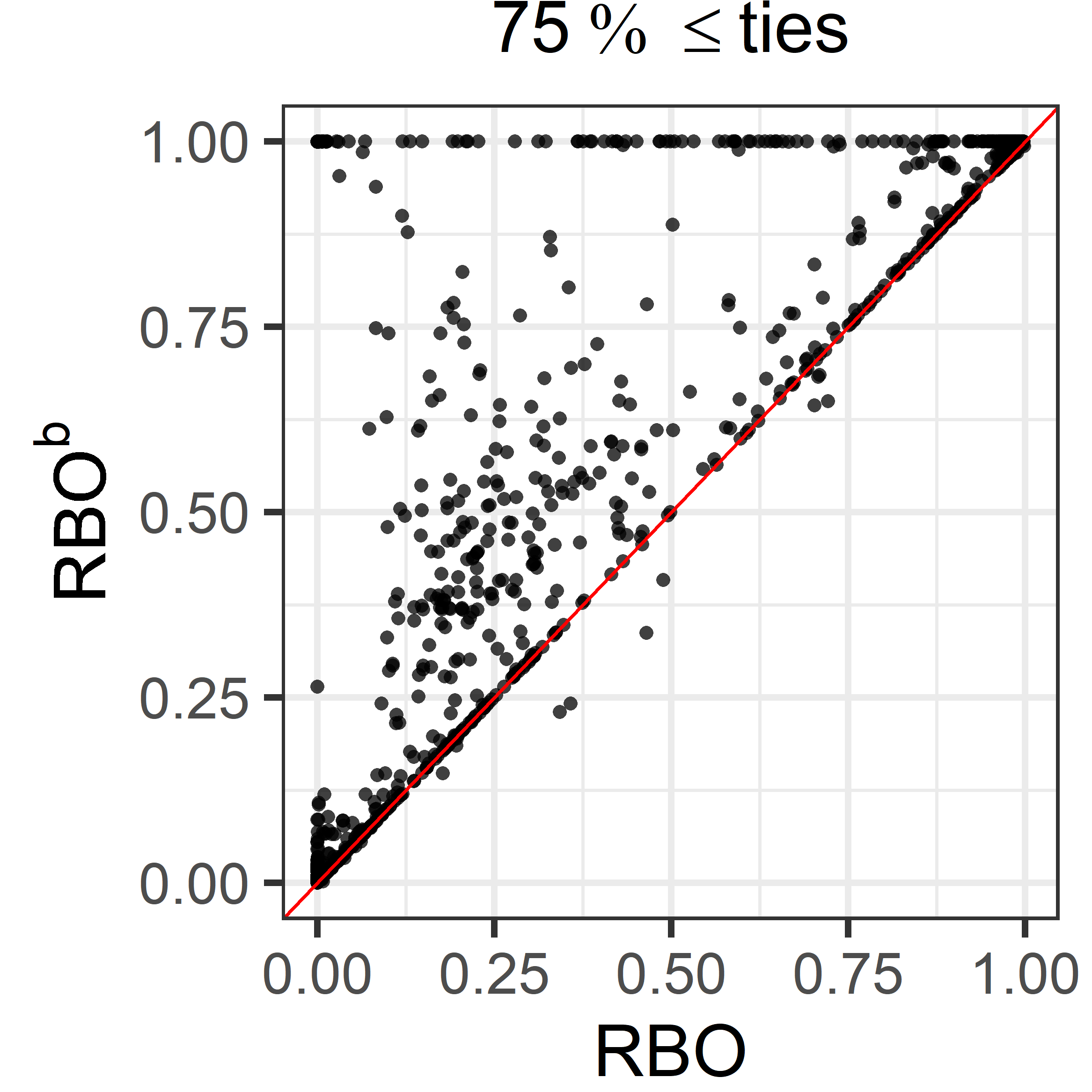}\\\vspace{2mm}
	\includegraphics[scale=\figscale]{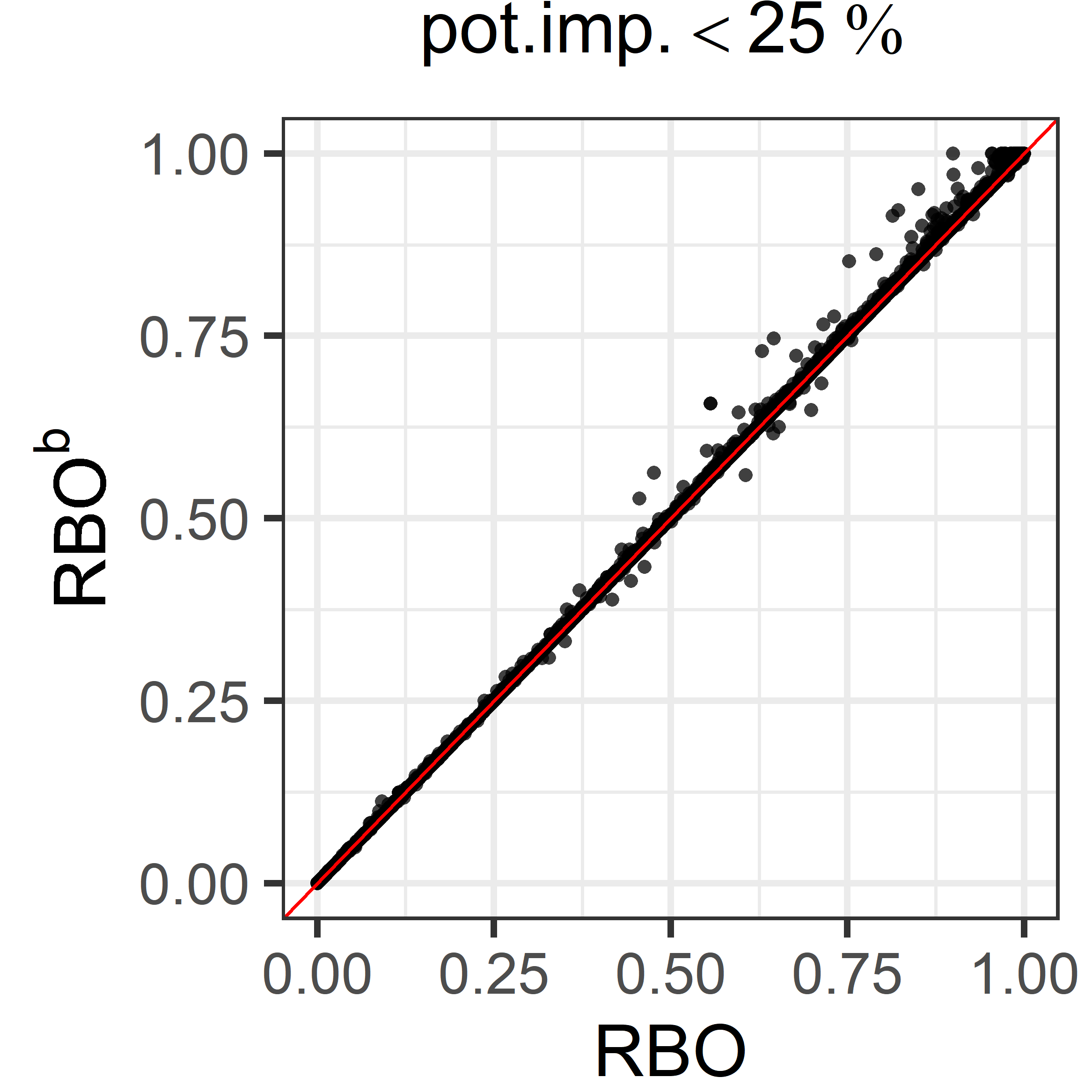}%
	\includegraphics[scale=\figscale]{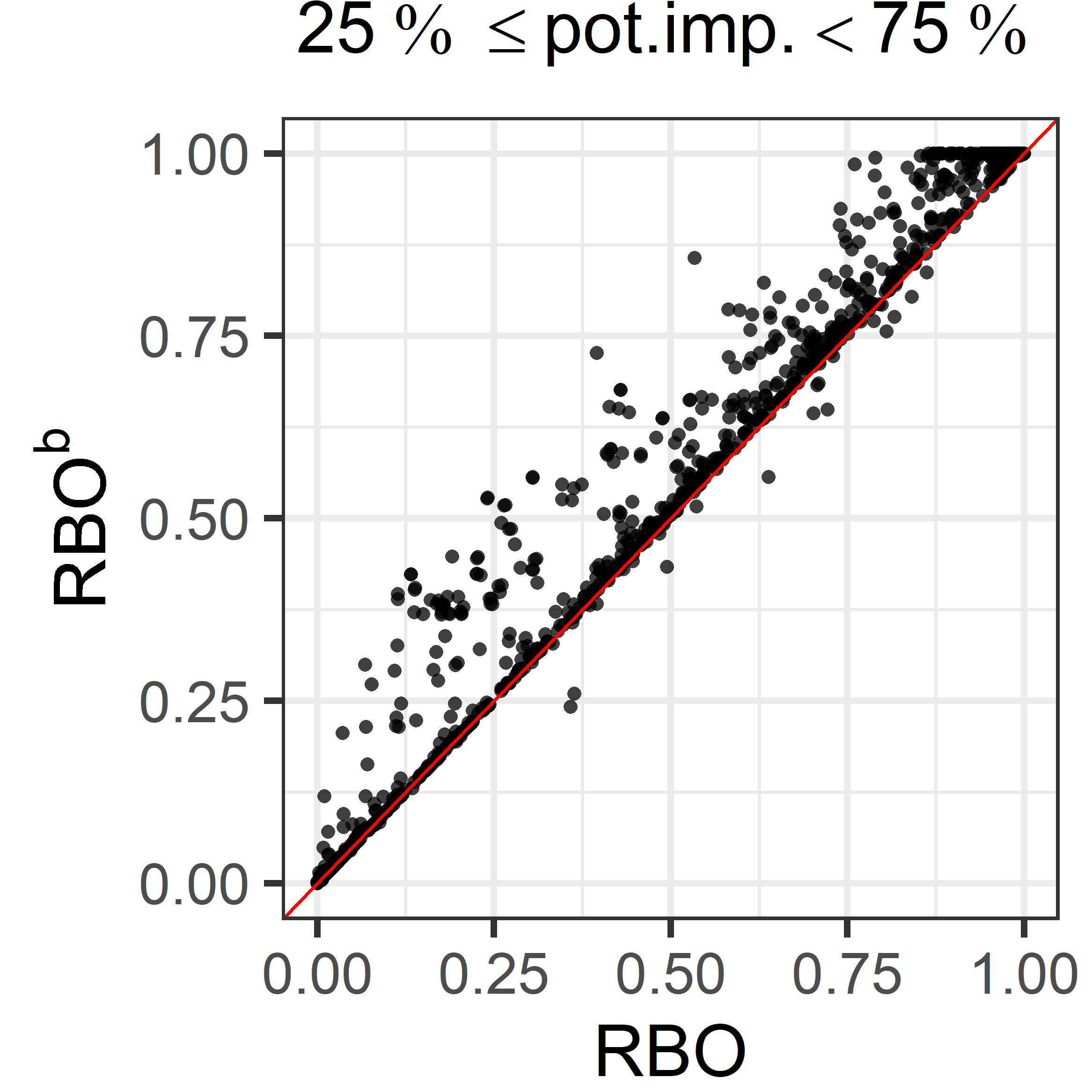}%
	\includegraphics[scale=\figscale]{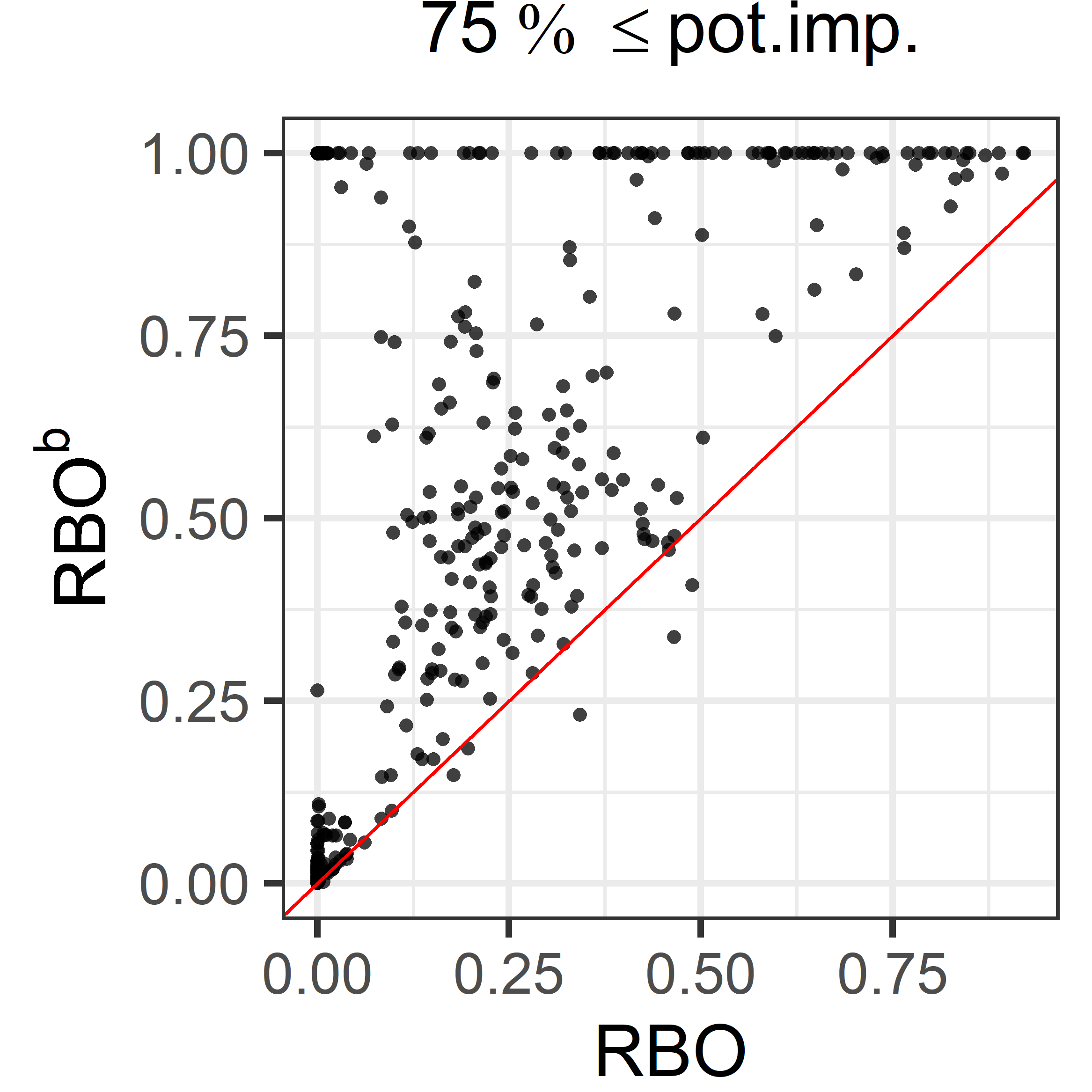}
	\caption{Differences between bare $\boldsec{RBO}$ (ties broken at random) and $\boldsec{RBO^b}$, faceted by the amount of tied items (top) and by their potential impact (bottom). TREC data, $\boldsec{p\!=\!0.9}$.}\label{fig:trec-nt}
\end{figure}

While these summary statistics give us a broad idea of the contrast between dealing with ties or not, we must note that the largest differences are mostly found between rankings with an extreme structure in one or two specific aspects.
The first aspect is the \emph{amount} of ties: $RBO^{\!w}$ and $RBO^b$ achieve maximum overlap when the tie groups are the same in both rankings, and the chance of this happening increases when all or most items are actually tied (e.g., 9\% of the rankings have at least 90\% of their items tied). For instance, Figure~\ref{fig:trec-nt}-top illustrates the case of bare $RBO$ vs $RBO^b$, faceting by the amount of ties found in the rankings. The most extreme differences indeed appear when most items are tied, although even with a small number of ties we can observe differences larger than 0.2.
The second aspect is the \emph{position} of the tied items: it is not enough to have many ties; they need to appear towards the top of the ranking in order to have an impact in the score. We may roughly quantify the potential impact of ties by simply summing the $p$-dependent weight of their ranks: $\big(\sum_d \ind{d~\text{is tied}}p^d\big) / \sum_d p^d$. Figure~\ref{fig:trec-nt}-bottom confirms that differences between bare $RBO$ and $RBO^b$ are more pronounced when the potential impact of ties is large. Therefore, rankings with a moderate number of ties may still exhibit large variations if those ties appear towards the top.

In general, Figure~\ref{fig:trec-nt} shows that the presence of ties does not immediately lead to a difference with respect to bare $RBO$, as this ultimately depends on the arrangement of those tied items. What ties give is room for these variations to be large. In other words, ties affect mostly the variance between measures, not the bias.

\subsection{Synthetic Data}\label{ssec:synthetic}

The results from the previous section should not generalize well to non-IR settings, as they involved quite long rankings, mostly even, and where the underlying domains (i.e. the document collections) are several orders of magnitude larger than the rankings, thus leading to a high degree of non-conjointness. 
In order to provide more general results, in this section we consider a synthetic dataset, generated as follows. Two rankings are generated with a Kendall $\tau$ between 0.5 and 1 over the same 1,000 items. Ties are introduced at random in each ranking and independent of the other, for a target tiedness between 10\% and 100\%, after which it is truncated to a length between 10 and 100. This was repeated 100,000 times, resulting in rankings with an average of 55 items, an average length difference of 30 items, and an average of 54\% items tied. In this case, we only break ties at random when computing bare $RBO$.

\begin{figure}[!t]
	\includegraphics[scale=\figscale]{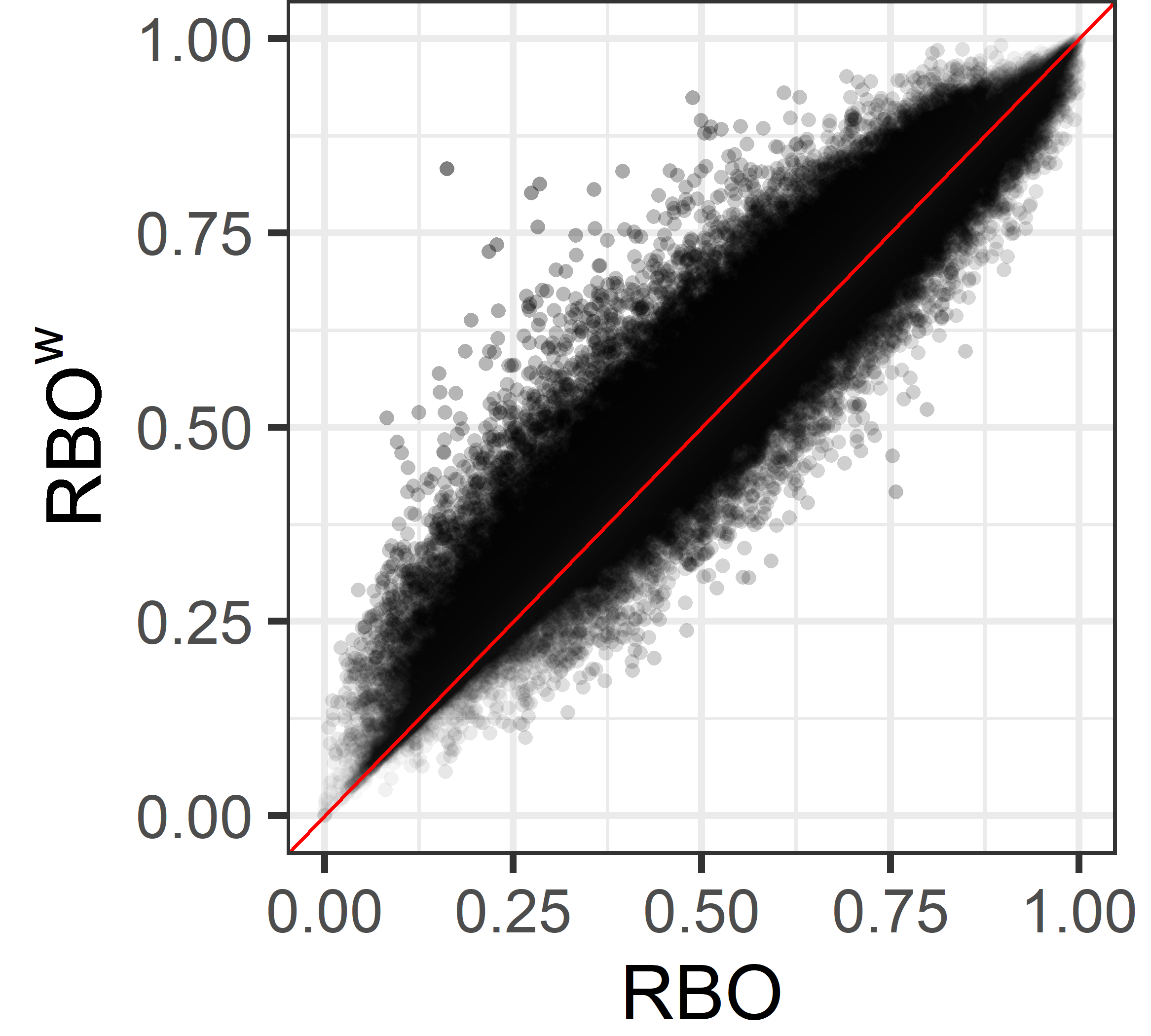}%
	\includegraphics[scale=\figscale]{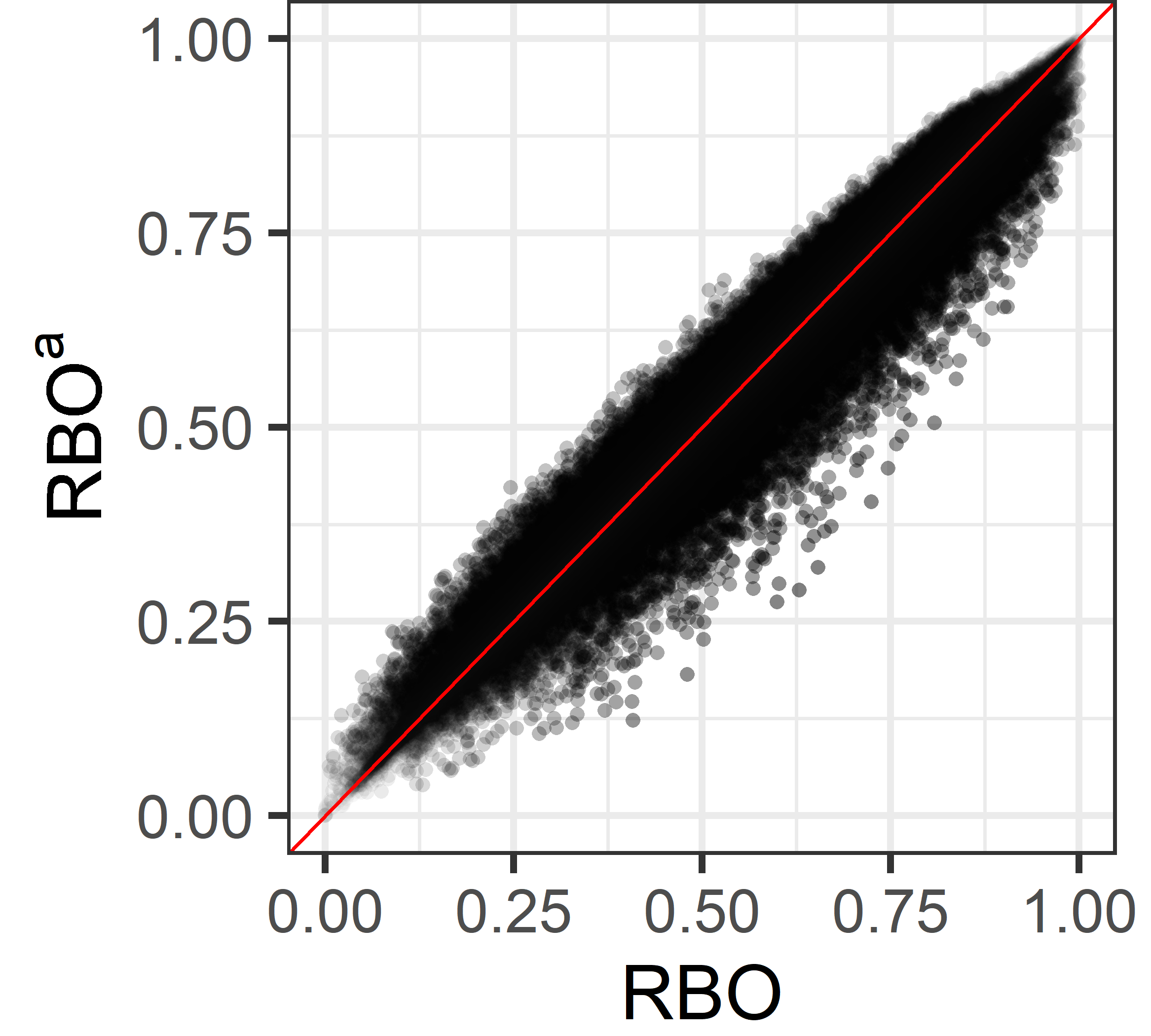}%
	\includegraphics[scale=\figscale]{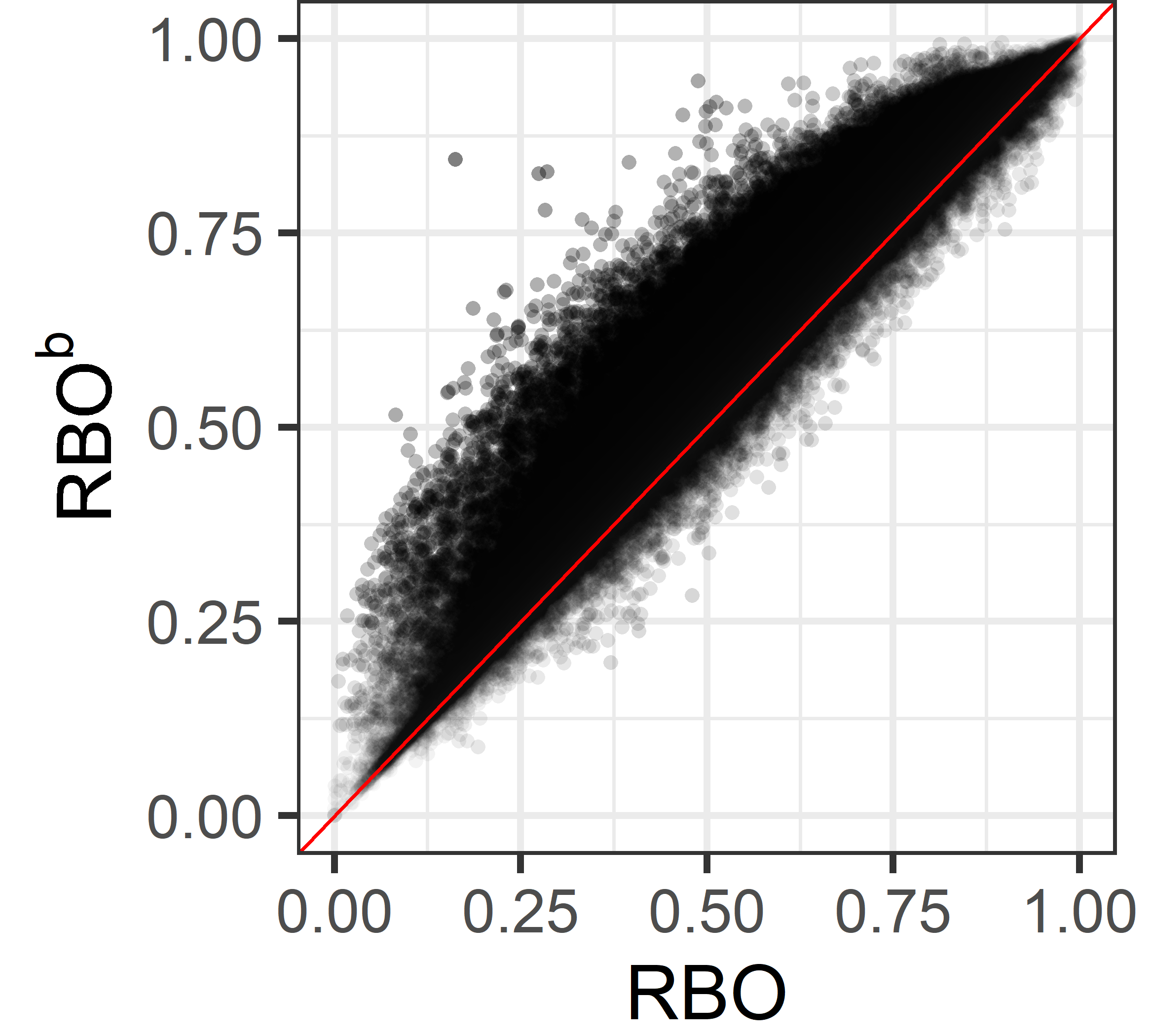}
	\caption{Differences between bare $\boldsec{RBO}$ (ties broken at random) and the three tie-aware variants. Synthetic data, $\boldsec{p\!=\!0.9}$.}\label{fig:synthetic}
\end{figure}

Figure~\ref{fig:synthetic} similarly compares all $RBO$ scores for $p=0.9$. In clear contrast with Figure~\ref{fig:trec}, we can first see that there are far fewer extreme deviations. This is because the simulated dataset does not contain rankings with such extreme structures as displayed in the TREC data.
In general, we see that $RBO^{\!w}$, and specially $RBO^b$, tend to produce higher scores than bare $RBO$, because at a given depth they allow all items in a group to contribute to the final score, not only those that are active at that depth. On the other hand, $RBO^a$ calculates the expected $RBO$ over permutations of the tied items, so differences are again nicely distributed around the diagonal.

Even if there are no extreme cases, deviations are generally larger than observed with the TREC data. As summarized in Table~\ref{tab:synthetic}, average deviations are 3 to 4 times larger, but most importantly, the amount of medium and large deviations is even an order of magnitude larger. In this case, the strategy followed to deal with ties brings a substantial difference in about 70\% of the comparisons.

\begin{table}[!t]
	\caption{Summary of differences between bare $\boldsec{RBO}$ and each of the three tie-aware variants. M for medium differences in $\boldsec{(0.01,0.1]}$, and L for large in $\boldsec{(0.1,1]}$. Synthetic data.}\label{tab:synthetic}
	\centering\small\setlength\tabcolsep{.75mm}\begin{tabular}{|r|rrrr|rrrr|rrrr|}
		\hline
		& \multicolumn{4}{c|}{$\left|RBO-RBO\!^w\right|$} & \multicolumn{4}{c|}{$\left|RBO-RBO^a\right|$} & \multicolumn{4}{c|}{$\left|RBO-RBO^b\right|$} \\ 
		$p$ & Avg. & Max. & M & L & Avg. & Max. & M & L & Avg. & Max. & M & L \\
		\hline
		0.8 & 0.07 & 0.76 & 50\% & 26\% & 0.05 & 0.51 & 52\% & 17\% & 0.08 & 0.77 & 46\% & 31\% \\
		0.9 & 0.04 & 0.67 & 64\% & 10\% & 0.03 & 0.34 & 63\% & 4\% & 0.06 & 0.68 & 56\% & 20\% \\
		0.95 & 0.03 & 0.53 & 63\% & 3\% & 0.02 & 0.25 & 56\% & $<$.01\% & 0.04 & 0.54 & 62\% & 8\% \\
		\hline
	\end{tabular}
\end{table}

\section{Conclusion and Recommendations}\label{sec:conclusions}

In this work we delved into the problem of tied items in Rank-Biased Overlap. First, we argued that the existing approach is incomplete, for it is unclear how to apply it to compute $RBO_\sub{EXT}$ and its bounds. More importantly, we showed that the notion of ties behind this approach (i.e. the sports ranking) is very different from the one traditionally used in the Statistics literature (i.e. uncertainty as to the actual order), most notably in Kendall's $\tau$.
We therefore developed two other variants of $RBO$ to accommodate this traditional view on ties.
Through a general formulation for prefix evaluation of $RBO$, we also showed how to fully compute all three variants.

Filling this gap, researchers can now make a conscious and sensible decision when dealing with ties. Our \textbf{recommendations} are:
\begin{itemize}
	\item When a tie represents equality, so that tied items \emph{really} occur at the same rank, one should compute $\boldsec{RBO^{\!w}}$.
	\item When a tie represents uncertainty, so that it is not known which item appears first:
	\begin{itemize}
		\item Ties should not be broken deterministically, such as by doc ID, because it inflates $RBO$ scores.
		\item Ties should not be broken at random because it introduces noise. $\boldsec{RBO^a}$ should be used instead, as it precisely computes the expected $RBO$ when breaking ties at random.
		\item If the measured overlap should be corrected by the amount of information lost due to ties, $\boldsec{RBO^b}$ should be used. This ensures $RBO^b(X,X)=1$, and implies $RBO^a\leq RBO^b$.
	\end{itemize}
\end{itemize}

As future work, we will bound the uncertainty introduced by ties, similarly to how bounds are used to quantify the uncertainty due to unseen items.

\begin{acks}
	Work facilitated by computational resources of the Delft AI Cluster at TU Delft. We dedicate this work to Akira Toriyama.
\end{acks}

\bibliographystyle{ACM-Reference-Format}
\balance
\bibliography{biblio}

\end{document}